\documentclass[12pt]{article}
\pdfoutput=1

\usepackage{amsmath}
\usepackage{amsfonts}
\usepackage{amssymb}

\usepackage[
      colorlinks=true,
      linkcolor=blue,
      urlcolor=blue,
      filecolor=black,
      citecolor=red,
      pdfstartview=FitV,
      pdftitle={},
        pdfauthor={Simon A. Gentle, Michael Gutperle, Chrysostomos Marasinou},
        pdfsubject={},
        pdfkeywords={},
        pdfpagemode={},
        bookmarksopen=true
      ]{hyperref}

\usepackage{color}
\usepackage{graphicx}
\usepackage{sectsty}
\sectionfont{\large}

\renewcommand{\thefootnote}{\fnsymbol{footnote}}
\renewcommand{\thanks}[1]{\footnote{#1}}
\newcommand{\starttext}{
\setcounter{footnote}{0}
\renewcommand{\thefootnote}{\arabic{footnote}}}
\newcommand{\<}{\langle}
\renewcommand{\>}{\rangle}

\numberwithin{equation}{section}

\def\cA{{\cal A}}

\def\cF{{\cal F}}

\def\cO{{\cal O}}

\DeclareMathOperator{\diag}{diag}
\usepackage{ dsfont } %for identity matrix
\def\id{\mathds{1}}

\renewcommand{\Im}{\operatorname{Im}}

\long\def\symbolfootnote[#1]#2{\begingroup%
\def\thefootnote{\fnsymbol{footnote}}\footnote[#1]{#2}\endgroup}

\begin{document}
\setlength{\baselineskip}{18pt}

\starttext
\setcounter{footnote}{0}

\begin{center}

{\Large \bf  Holographic entanglement entropy of surface defects}

\vskip 0.4in

{\large  Simon A.\ Gentle, Michael Gutperle and  Chrysostomos Marasinou}

\vskip .2in

{ \it Department of Physics and Astronomy }\\
{\it University of California, Los Angeles, CA 90095, USA}\\[0.5cm]
\href{mailto:sgentle@physics.ucla.edu}{\texttt{sgentle}}\texttt{, }\href{mailto:gutperle@physics.ucla.edu}{\texttt{gutperle}}\texttt{, }\href{mailto:cmarasinou@physics.ucla.edu}{\texttt{cmarasinou@physics.ucla.edu}}

\end{center}

\bigskip

\bigskip
 
\begin{abstract}

\setlength{\baselineskip}{18pt}

We calculate the holographic entanglement entropy in type IIB supergravity solutions that are dual to half-BPS disorder-type surface defects in ${\cal N}=4$ supersymmetric Yang-Mills  theory.  The entanglement entropy  is calculated for  a ball-shaped region bisected by a surface defect. Using the bubbling supergravity solutions we also compute the expectation value of the defect operator. Combining our result with the previously-calculated one-point function of the stress tensor in the presence of the defect, we adapt  the calculation of Lewkowycz and Maldacena \cite{Lewkowycz:2013laa} to obtain a second expression for the entanglement entropy. Our two expressions agree up to an additional term, whose  possible origin and significance is discussed.

\end{abstract}

\setcounter{equation}{0}
\setcounter{footnote}{0}

\newpage

\tableofcontents

\newpage

%%%%%%%%%%%%%%%%%%%%%%%%%%%%
\section{Introduction}
\label{sec1}
%%%%%%%%%%%%%%%%%%%%%%%%%%%%

Entanglement entropy is an important quantity that measures the quantum entanglement 
between different regions of a system. It furnishes an order parameter for phase transitions and is central to the recent efforts to explore  the relation between   quantum entanglement and geometry. The Ryu-Takayanagi proposal   
  \cite{Ryu:2006bv,Ryu:2006ef} allows one to calculate  entanglement entropies in theories described by a holographic AdS/CFT dual.  

The simplest setup  for which entanglement entropy can be calculated is a spherical entangling surface in the ground state of a given theory.  In recent years many generalizations  have been studied both on the field theory side and via holography, including more general entangling  surfaces, time dependence, finite temperature and other systems  not in their ground state.
  
 Of particular interest is the entanglement entropy in the presence of  non-local operators. Two  types of non-local operator can be distinguished. Operators such as Wilson lines  can be expressed  as operator insertions written in terms of the fundamental fields of the theory.  However, disorder-type operators cannot be written in this way and are instead   characterized by   the singular behavior of the fundamental fields close to a defect. One example of the latter is the 't Hooft loop in gauge theories.  Note that $S$-dualities often map defects of the two types into each other \cite{Kapustin:2006pk}.

The entanglement entropy for co-dimension one Janus-like defects and boundary CFTs have been studied in   \cite{Azeyanagi:2007qj, Chiodaroli:2010mv, Chiodaroli:2010ur, Estes:2014hka, D'Hoker:2014dla, Gutperle:2015hcv}. 
%MG
In \cite{Jensen:2013lxa} the entanglement entropy in the presence of defects was related to a thermal entropy on a hyperbolic space, applying the methods of \cite{Casini:2011kv} to the presence of a defect.
%MG

 In  \cite{Lewkowycz:2013laa} the entanglement entropy in the presence of a (supersymmetric) Wilson line operator was calculated in four-dimensional ${\cal N}=4$ $SU(N)$  SYM theory as well as three-dimensional   ABJM theories.  In the holographically dual theories the description of the Wilson line operator depends on the size of the representation: for representations with Young tableau  of order $1$, $N$ and $N^2$  the Wilson line  is described by a fundamental string~\cite{Maldacena:1998im,Rey:1998ik}, 
 probe D-branes~\cite{Gomis:2006sb,Gomis:2006im,Yamaguchi:2006tq} and bubbling supergravity solutions~\cite{D'Hoker:2007fq}\footnote{See \cite{Yamaguchi:2006te,Lunin:2006xr} for earlier work on bubbling solutions dual to Wilson lines.}, respectively. In~\cite{Gentle:2014lva}  the 
 holographic entanglement entropy for the bubbling supergravity solution was computed and exact agreement between the field theory and holographic calculations was found.

Surface operators have received much less attention. In the present paper we focus on disorder-type surface defects  in four-dimensional ${\cal N}=4$ $U(N)$  SYM theory  constructed  in~\cite{Gukov:2006jk,Gukov:2008sn}.  Their dual description as bubbling geometries of type IIB supergravity was identified in \cite{Gomis:2007fi} using the  solutions constructed in~\cite{Lin:2004nb,Lin:2005nh}.  For notational ease we will drop the qualifier `disorder-type' and simply call these `surface defects'.

\begin{figure}[!t]
  \centering
  \includegraphics[width=63mm]{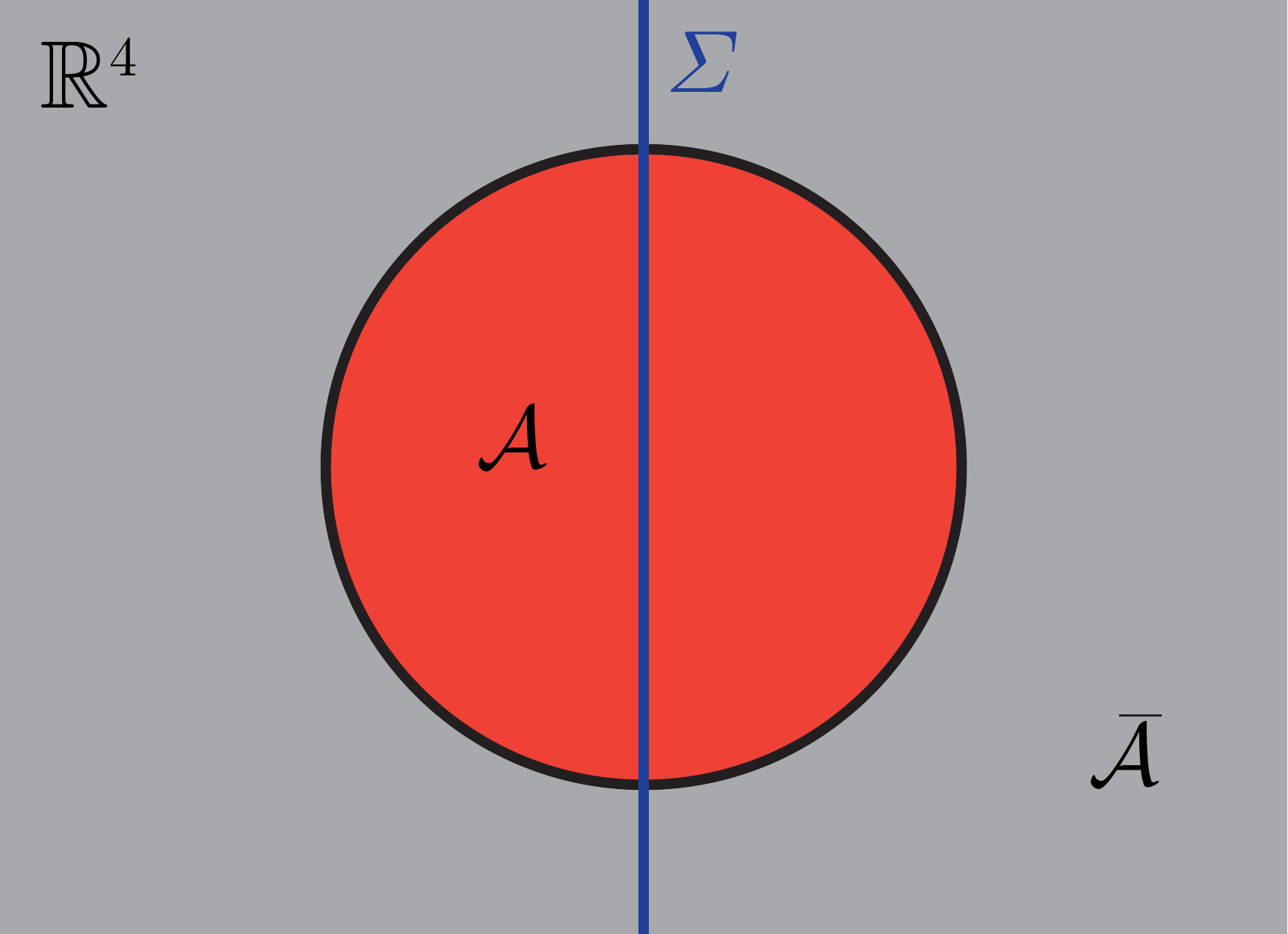}
  \caption{Geometry of the entangling region and surface defect in $\mathbb{R}^{4}$. The entangling region ${\cal A}$ is a three-dimensional ball with a two-sphere boundary. The surface defect extends along a spatial line and bisects the two-sphere.}
 \label{fig:R4-setup}
\end{figure}

 The geometric setup 
of the surface defect is best visualized in $\mathbb{R}^{4}$. At fixed time  the entangling region  ${\cal A}$ is a three-dimensional ball with a spherical boundary. The surface defect $\Sigma$ 
extends in one spatial direction (and time). We depict the setup in figure \ref{fig:R4-setup}, with 
one spatial and the time direction suppressed. Note that unlike the case of the Wilson line, it is 
generic for the surface defect and the boundary of the entangling space to intersect. In 
the text we also use different geometries, namely $AdS_3\times S^1$ and $S^1\times H^3$, which are related to $\mathbb{R}^4$ by  a coordinate change and Weyl rescaling. 

The goal of this paper is to calculate the holographic entanglement entropy  in the presence of  surface defects   for ${{\cal N}=4}$ SYM and compare them to the result  obtained by mapping the entanglement entropy to a thermal entropy as in \cite{Lewkowycz:2013nqa}. Here we 
calculate  the expectation values  of the surface defect  using holography and use the 
expectation value of the stress tensor that was previously obtained in \cite{Drukker:2008wr}. The 
methods used in this paper closely follow those used in our previous paper \cite{Gentle:2015jma} which addressed the same questions for  Wilson surface operators  \cite{Ganor:1996nf,Berenstein:1998ij,Corrado:1999pi} in six-dimensional $(2,0)$ theory  \cite{Witten:1995zh,Strominger:1995ac} and their dual supergravity solutions \cite{D'Hoker:2008wc,D'Hoker:2008qm}.

The structure of this paper is as follows. In section \ref{sec2} we  review the field theory description of half-BPS surface defects in ${{\cal N}=4}$ SYM theory. In section \ref{sec3} we  review the bubbling supergravity solutions  dual to these defects. In section \ref{sec4} we calculate the entanglement entropy for a spherical entangling region that intersects the surface defect. In section \ref{sec5} we calculate the expectation value of the surface defect by evaluating the on-shell supergravity action on the bubbling solution and review the result for the one-point function of the stress energy tensor in the presence of a surface defect. In section \ref{sec6} the expectation values are used  to calculate the entanglement entropy following the method of Lewkowycz and Maldacena~\cite{Lewkowycz:2013nqa} which we then compare with our  holographic result. The two entanglement entropies do not match completely and we discuss  possible explanations  for the mismatch in section \ref{sec7}. Various technical details are presented in the appendices.

%%%%%%%%%%%%%%%%%%%%%%%%%%
\section{Review of surface defects in ${\cal N}=4$ SYM}
\setcounter{equation}{0}
\label{sec2}
%%%%%%%%%%%%%%%%%%%%%%%%%%%%

In this section we  review the construction of half-BPS surface defects in ${\cal N}=4$ SYM theories first obtained in \cite{Gukov:2006jk} and studied in detail holographically in \cite{Drukker:2008wr,Gomis:2007fi}. 

The defects are supported on a two-dimensional surface $\Sigma$ in $\mathbb{R}^4$.  They are disorder-type operators so, unlike Wilson line operators, cannot be written as an 
integral of the fundamental gauge fields over $\Sigma$. Instead, they are characterized by singularities of the gauge fields and/or  scalar fields at the surface $\Sigma$ as 
well as holonomies along cycles in the space normal to the surface. Furthermore, we are 
interested in  half-BPS defects that preserve half the superconformal symmetry $PSU(1,1|
2)\times PSU(1,1|2) $ inside $ PSU(2,2|4)$. For such superconformal defects it is possible to 
perform a Weyl transformation from $\mathbb{R}^4$ to $AdS_3\times S^1$, in which the surface $\Sigma$ is 
mapped to the boundary of $AdS_3$. This has two advantages: first, the singularities of the 
fields along $\Sigma$ are mapped to boundary behavior in $AdS_3$ and second, the 
$AdS_3\times S^1$ geometry appears naturally in the dual bubbling supergravity solutions that we will review in section \ref{sec3}.

The half-BPS surface defect is characterized by the following data. The  non-trivial conditions on the gauge field and scalars break the $U(N)$ gauge group to the Levi subgroup $L=\prod_{i=1}^M U(N_i)$ with $M$ factors. Near the boundary of $AdS_3$ the gauge field has a non-vanishing component along the $U(1)$ coordinate, which we denote by $\psi$:
\begin{equation}\label{aadsf}
A_ \psi = \diag\left\{\alpha_1\id_{N_1},\alpha_2\id_{N_2},\ldots,\alpha_M\id_{N_M}\right\}\quad \textrm{with}\quad \sum_{i=1}^M N_i = N
\end{equation}
There are $M$ theta angles for the $M$ unbroken $U(1)$ factors (see \cite{Gomis:2007fi,Drukker:2008wr} for details), which can be parametrized by the matrix
\begin{equation}
\eta = \diag\left\{\eta_1\id_{N_1},\eta_2\id_{N_2},\ldots,\eta_M\id_{N_M}\right\}
\end{equation}
A complex scalar, which we can choose as $\Phi=\phi_5+i \phi_6$,   has non-trivial behavior along the $S^1$:
\begin{equation}\label{phadsf}
\Phi ={e^{-i\psi}\over \sqrt{2}}\, \diag\left\{\left( \beta_1+i \gamma_1\right)\id_{N_1},\left( \beta_2+i \gamma_2\right)\id_{N_2},\ldots,\left( \beta_M+i \gamma_M\right)\id_{N_M}\right\}
\end{equation}
To summarize, the surface defect is characterized by the set of $M$ integers $N_i$ and a set of $4M$ real parameters $(\alpha_i,\eta_i,\beta_i,\gamma_i)$ with $i=1,2,\ldots,M$.  

We  also cite the results for the expectation value of the surface defect and the one-point function of the stress tensor calculated in \cite{Drukker:2008wr}  in order to compare them with the results of our holographic calculations. In the semiclassical approximation the expectation value of the surface operator is determined by evaluating the classical ${\cal} N=4$ SYM action on the field background. It was shown in \cite{Drukker:2008wr} that this gives zero and hence
\begin{align}
\langle {\cal O}_\Sigma\rangle&= e^{-S_{YM}}|_{\rm surface} =1
\end{align}
In addition, several one-point functions of local operators and Wilson line operators in the presence of the surface defect were calculated in \cite{Drukker:2008wr}. The only one which is relevant for the present paper is the one-point function of the stress tensor, which takes the following form due to $AdS_3\times S^1$ symmetry and the fact that the stress tensor is traceless:
\begin{equation}\label{stressTensorForm}
\langle T_{\mu\nu}\rangle_\Sigma\, dx^\mu\, dx^\nu = h_\Sigma \left(ds_{AdS_3}^2 - 3\, d\psi^2\right)
\end{equation}
The semiclassical value for the scaling weight $h_\Sigma$ 
is found by evaluating the stress tensor of ${\cal N}=4$ SYM on the field background:
\begin{equation}
h_\Sigma = -{2\over 3 g_{YM}^2} \sum_{i=1}^M N_i (\beta_i^2+\gamma_i^2)
\end{equation}

%%%%%%%%%%%%%%%%%%%%%%%%%%%%
\section{Review of bubbling supergravity solutions}
\setcounter{equation}{0}
\label{sec3}
%%%%%%%%%%%%%%%%%%%%%%%%%%%%

In \cite{Drukker:2008wr,Gomis:2007fi} it  was proposed  that the solution found in \cite{Lin:2004nb,Lin:2005nh} is the holographic dual of the surface defect operator. The solution is constructed as a $AdS_3\times S^3\times U(1)$ fibration over a three-dimensional space  with boundary parametrized by the coordinates $y,x_1,x_2$, where the boundary is located at $y=0$. The metric takes the form
\begin{equation}\label{metans}
ds^2 = y \sqrt{\frac{2 f + 1}{2 f - 1}}\, ds^2_{AdS_3} + y \sqrt{ \frac{2 f - 1}{2 f + 1} }\, ds^2_{S^3} + \frac{2 y}{\sqrt{4f^2-1}} \left( d\chi + V \right)^2 +  \frac{\sqrt{4f^2-1}}{2 y}\, ds^2_X
\end{equation}
where the $AdS_3$ metric is in Poincar\'{e} coordinates and the metric on the base is simply the flat Euclidean metric:
\begin{equation}\label{basemetric}
ds_{AdS_3}^2={dt^2+dl^2+ dz^2\over z^2}\quad\textrm{and}\quad ds^2_X=(dy^2+dx_1^2+dx_2^2)
\end{equation}
The function $f(y,x_1,x_2)$ satisfies a linear partial differential equation with $M$ sources located in the bulk of the base space $X$ at $y=y_i, x=\vec{x}_i$ with $ i=1,2,\ldots,M$:
 \begin{align}\label{fdiffeq}
 \partial_1^2 f+\partial_2^2 f+y \partial_y \left( \frac{\partial_y f}{y}\right)  = \sum_{i=1}^M 2 \pi y_i\, \delta(y-y_i)\, \delta^2(\vec{x} -\vec{x}_i)
\end{align}
$V$  is a one-form on $X$ that can be obtained from $f$ by solving
\begin{align}\label{vdeterm}
dV = {1\over y} \star_3 d f
\end{align}
Note that (\ref{vdeterm}) only fixes $V$ up to an exact  form and the freedom to 
redefine $V\to V+d\omega$ will be important to obtain a  manifestly asymptotically $AdS$ metric as detailed in appendix~\ref{app:FGCoordinates:GaugeChoice}.
The only other non-trivial field is the self-dual five-form field strength, which takes the form
\begin{align}\label{F5Solution}
F_5 =&- {1\over 4} \Big( d\Big[ y^2{2f+1\over 2f-1}(d\chi +V) \big]- y^3 \star_3 d\Big[ {f+1/2\over y^2}\Big]\Big) \wedge \omega_{AdS_3} \nonumber\\
 &\;\;- {1\over 4} \Big( d\Big[ y^2{2f-1\over 2f+1}(d\chi +V) \big]- y^3 \star_3 d\Big[ {f-1/2\over y^2}\Big]\Big) \wedge \omega_{S_3}
 \end{align}
 where $\star_3$ denotes the Hodge dual\footnote{The sign of the Hodge dual is fixed by $\star_3 dy =dx_1\wedge dx_2$ and cyclic permutations of $dy,dx_1$ and $dx_2$.}  in the-three dimensional base space $X$ with metric given by (\ref{basemetric}).

This solution was first constructed in \cite{Lin:2005nh} as a double analytic continuation of the LLM solution \cite{Lin:2004nb}. Indeed (\ref{metans}) becomes the LLM metric by continuing the $U(1)$ fiber coordinate to a time like coordinate and continuing $AdS_3$ to $S^3$. Note however that the boundary condition on the function $f$ is different:  the $AdS_3$ volume can never shrink to zero size in a smooth solution so we must have $f\to {\small 1\over2 }$ as $y$ approaches the boundary of $X$. Hence   for the bubbling surface solution the coloring of the boundary determined by the regions where $\lim_{y\to 0}f=\pm {\small 1\over2 }$ in the LLM solution gets replaced by the bulk sources in (\ref{fdiffeq}).

\medskip

The supergravity solutions depend on $3M$ parameters, which are the $M$ sources  on the right hand side of (\ref{fdiffeq}),  located in $X$   at $y_i, \vec{x}_i$ with $ i=1,2,\ldots,M$. There is an overall translation symmetry along $\vec{x}$; this allows us to choose `center-of-mass' coordinates, which sets 
\begin{equation}\label{eq:CM}
\vec{x}^{(0)} \equiv \sum_{i=1}^M y_i^2 \vec{x}_i =0
\end{equation} 
This choice will make the expressions considerably more compact.
The general solution of (\ref{fdiffeq})  for the function  $f$  is then given by
\begin{equation}\label{fforma}
f = \frac{1}{2} + \sum_{i=1}^M f_i
\end{equation}
with
\begin{equation}\label{eq:fiformula}
f_i = -\frac{1}{2} + \frac{(\vec{x}-\vec{x}_i)^2+y^2+y^2_i}{2\sqrt{[(\vec{x}-\vec{x}_i)^2+y^2+y^2_i]^2-4y^2 y^2_i}}
\end{equation}
For such an $f$ the solution of  the differential equation (\ref{vdeterm}) for the one-form $V$ is given by
\begin{align}
V_I\, dx^I &= -\sum_{i=1}^M  \sum_{I,J } \epsilon_{IJ} \frac{(x_J-x_{iJ})[(\vec{x}-\vec{x}_i)^2+y^2-y^2_i]}{2(\vec{x}-\vec{x}_i)^2\sqrt{[(\vec{x}-\vec{x}_i)^2+y^2+y^2_i]^2-4y^2 y^2_i}} \, dx^I \label{eq:Vdefinition}
\end{align}
where the indices $I,J$ run over $x_1,x_2$.  

In \cite{Drukker:2008wr,Gomis:2007fi} the parameters of the supergravity solution were identified with the parameters of the gauge theory surface defect as follows:
\begin{equation}\label{fieldTheoryToSugra}
{1\over 2\pi l_s^2}\left(x_{i1}+ i x_{i2}\right) = \beta_i+ i\gamma_i,\quad \frac{y_i^2}{L^4} = {N_i\over N}
\end{equation}
where $L$ denotes the radius of  $AdS_5$.  The parameters $\alpha_i$ and $\eta_i$ are identified with periods of the NSNS and RR two-form potentials on non-trivial  two-cycles in the solutions. On the supergravity side these periods carry  only topological information since the three-form field strengths of the two-form potentials vanish. As the calculations performed  in section~\ref{sec4} and \ref{sec5} depend only on the metric and the five-form,  we conclude that all our calculations will be independent  of the periods and hence the parameters  $\alpha_i$ and $\eta_i$.

\subsection{The vacuum solution}
In order to develop intuition for the geometry it is useful to consider the 
 $AdS_5\times S^5$ vacuum solution, which can be obtained by considering only one source,  i.e.\ setting  $M=1$. Translation invariance allows one to set   $\vec{x}_1=0$ and from \eqref{fieldTheoryToSugra} we can fix $y_1=L$ since $N_1=N$. To exhibit the  $AdS_5\times S^5$ metric explicitly it is convenient to introduce new coordinates:
\begin{align} 
 \chi &= \frac{1}{2} \left(\psi - \phi\right)  \nonumber\\
   y &= L^2\sqrt{\rho^2+1}\, \cos\theta \nonumber \\
  x_1 &= x_1^{(0)}+  L^2 \rho \sin\theta \cos\left(\psi + \phi\right) \nonumber\\
  x_2 &= x_2^{(0)} + L^2 \rho \sin\theta \sin\left(\psi + \phi\right) \label{eq:changecoordinates}
\end{align}
where the range of the angular variables is given by $\theta\in[0,\pi/2]$, $\psi\in [0,2\pi]$, $\phi\in [0,2\pi]$. It is straightforward to verify that for this choice the 
function $f$ (\ref{fforma}) and the one-form $V$ (\ref{eq:Vdefinition})  for  the vacuum solution take the following form
\begin{equation}\label{eq:vacuumfandV}
f = \frac{1}{2}\,  \frac{\rho^2 + \cos^2\theta +1}{\rho^2+ \sin^2\theta} \quad \textrm{and} \quad 
V = \frac{1}{2}\,  \frac{\rho^2 - \sin^2\theta}{\rho^2+ \sin^2\theta}\, d( \psi + \phi)
\end{equation}
where the gauge transformation can be set to zero, i.e.  $\omega = 0$. Using the expressions  given in (\ref{metans}) the metric  can be calculated and gives
\begin{equation}\label{eq:AdSmetric}
ds^2 = L^2\left[ \left(\rho^2+1\right) ds^2_{AdS_3} + \frac{d\rho^2}{\rho^2+1} + \rho^2\, d\psi^2  + d\theta^2 + \sin^2\theta\, d\phi^2 + \cos^2\theta\, ds^2_{S^3} \right]
\end{equation}
which is indeed $AdS_5\times S^5$.  Note that the metric is written in a form for which the conformal boundary is $AdS_3\times S^1$. In the following we will set the $AdS$ radius $L=1$ and restore it by dimensional analysis when needed.

\subsection{Asymptotics and regularization of the bubbling solution}
The integrals appearing later in the holographic entanglement entropy and the expectation value calculations are divergent. Therefore, we need to regulate them introducing a cut-off. In this section we map the general metric to a Fefferman-Graham (FG) form \eqref{eq:FGmetric}, we find the FG coordinate map \eqref{eq:FGMap} and derive the cut-off surface \eqref{eq:cut-off} in terms of the FG UV cut-off. 

The fact that a general solution must be asymptotically $AdS_5 \times S^5$ implies the following restriction on $y_i$:
\begin{equation}\label{eq:AdSradius}
\sum_{i=1}^M y_i^2=1
\end{equation}
It is straightforward to see this considering the map to field theory parameters \eqref{fieldTheoryToSugra} and $ \sum_{i=1}^M N_i=N$.

We will work in the coordinate system introduced in \eqref{eq:changecoordinates} and expand the general solution at large $\rho$. As mentioned above, the one-form $V$ is defined up to an exact form. Thus, we use a gauge transformation to remove the $V_\rho$ component of this vector. This brings the metric into a manifestly asymptotic form and makes it as compact as possible which is convenient for our calculations. Fixing the gauge, $\omega$  becomes:
\begin{equation}\label{omegaResultImplicit}
\omega  =- \frac{M-1}{2}\, \alpha  +  \sin\theta  \sum_{n=1}^{\infty} \frac{ V_1^{(n+1)}(\theta,\alpha)\cos\alpha +  V_2^{(n+1)}(\theta,\alpha)\sin\alpha}{n\rho^n} \\
\end{equation}
where the $V_I^{(n+1)}$ are the coefficients in a large $\rho$ expansion of the functions given in \eqref{eq:Vdefinition}. The detailed procedure and the explicit form of $\omega$ are given in the appendix \ref{app:FGCoordinates:GaugeChoice}. 

The next step is to write the metric in terms of the $\{\rho,\psi,\theta,\phi\}$ coordinates. We write it as a deviation of the vacuum \eqref{eq:AdSmetric}:
\begin{align}\label{largeRhoMetric}
ds^2 &= \frac{1}{\left(\rho^2+1\right)} \left( 1+ F_\rho\right) d\rho^2+ \left(\rho^2+1\right) \left(1+F_1\right)\, ds^2_{AdS_3} + \rho^2\left( 1+ F_2\right) d\psi^2 \nonumber \\
&\phantom{=\ } + \cos^2\theta \left(1+F_3\right)\, ds^2_{S^3} + \left( 1+ F_4\right) d\theta^2 + \sin^2\theta \left( 1+ F_5\right) d\phi^2  \nonumber \\
&\phantom{=\ }  +F_6\, d\theta\, d\psi  +F_7\, d\psi\, d\phi +F_8\, d\theta\, d\phi  
\end{align}
with the $F_a$ being functions of $\{\rho,\theta,\alpha \equiv \psi+ \phi\}$ expanded at large $\rho$. Specifically, $F_\rho, F_m \sim O\left(\rho^{-2}\right)$ for $m\in \{1,2,\ldots,7\}$ and $F_{8} \sim O\left(\rho^{-4}\right)$. Only certain coefficients in the $F_\rho$ expansion emerge in our calculations and their expressions are given in the appendix \ref{app:FGCoordinates:CoordinateMap}. These coefficients are expressed in terms of dimensionless moments. We will mainly express quantities in terms of these moments throughout the paper and therefore it is convenient to define them in advance:
\begin{equation}\label{momentsDefinition}
m_{abc} \equiv \sum_{i=1}^M y_i^a x_{i1}^b x_{i2}^c
\end{equation}
Note that for the $AdS_5\times S^5$ vacuum only the following moments are non-zero:
\begin{equation}\label{eq:vacmoments}
m^{ (0)}_{k00}=1 \quad\textrm{for}\quad  k=2,4,6,\ldots
\end{equation}

A general bubbling solution, preserving the $AdS_3 \times S^3 \times S^1$ isometry, can then be written in the following Fefferman-Graham form:
\begin{align}\label{eq:FGmetric}
ds^2 &= \frac{1}{u^2} \left(du^2 +\alpha_1\, ds^2_{AdS_3} + \alpha_2\, d\tilde\psi^2\right)  + \alpha_3\, ds^2_{S^3} + \alpha_4\, d\tilde{\theta}^2 + \alpha_5\, d\tilde{\phi}^2  \nonumber \\ 
&\phantom{=\ }+\alpha_6\, d\tilde{\theta}\, d\tilde{\psi} +\alpha_7\, d\tilde{\psi}\, d\tilde{\phi} +\alpha_8\, d\tilde{\theta}\, d\tilde{\phi} 
\end{align}
The condition that the metric must asymptote to $AdS_5\times S^5$ with $AdS_3\times S^1$ boundary implies that the new coordinates $u,\tilde\psi,\tilde\theta,\tilde\phi$ and the $\alpha_m$ (expressed as functions of $\rho,\psi,\theta,\phi$) fall off as 
\begin{gather}
u = \frac{1}{\rho}\left( 1+ \ldots \right), \quad \tilde\psi = \psi + \ldots, \quad \tilde\theta =  \theta + \ldots , \quad \tilde\phi = \phi + \ldots \nonumber\\
\alpha_1= 1+ \ldots, \quad \alpha_2 =  1 + \ldots ,\quad \alpha_3 = \cos^2\theta\left(1 + \ldots\right),\quad \alpha_4 = 1 + \ldots \nonumber \\
 \alpha_5 = \sin^2\theta\left(1 + \ldots\right), \quad \alpha_6 =   \ldots ,\quad \alpha_7 =  \ldots, \quad \alpha_8 = \ldots \nonumber
\end{gather}
The ellipses denote powers of $\rho^{-1}$ whose coefficients are determined by equating \eqref{largeRhoMetric} and \eqref{eq:FGmetric}. The explicit coordinate map is given in \eqref{eq:FGMap}.

The integrals in the entanglement entropy and expectation value calculations diverge at large $\rho$. It is useful to express the coordinate map as a cut-off relation $\rho = \rho_c(\varepsilon,\psi,\theta,\phi)$. This is found by solving the first equation in \eqref{eq:FGMap} for $\rho$ at the small $u$ limit and identifying $u$ with the FG cut-off, $u=\varepsilon$. The outcome is:
\begin{align}
\rho_c(\varepsilon,\psi,\theta,\phi) &=  \frac{1}{\varepsilon} + \frac{F_\rho^{(2)}-1}{4}\,\varepsilon +  \frac{F_\rho^{(3)}}{6}\, \varepsilon^2   \label{eq:cut-off}  \\
&\phantom{=\ } + \frac{16\left[F_\rho^{(4)}-F_\rho^{(2)}\left(F_\rho^{(2)}-1\right)\right]-\left(\partial_\theta F_\rho^{(2)}\right)^2- \left(\partial_\phi F_\rho^{(2)}\right)^2 \csc^2\theta}{128}\,  \varepsilon^3 +O\left(\varepsilon^4\right)\nonumber
\end{align}
Once we substitute for the coefficients of $F_\rho$ we find that this function can be written as $\rho_c(\varepsilon,\theta,\alpha)$ with $\alpha = \psi+ \phi$.

%%%%%%%%%%%%%
\section{Holographic entanglement entropy}
\label{sec4}
%%%%%%%%%%%%%

The Ryu-Takayanagi prescription \cite{Ryu:2006bv,Ryu:2006ef} states that the entanglement entropy of  a spatial region ${\cal A}$  is given by the area of a co-dimension two  minimal surface $\cal M$ in the bulk that is anchored on the $AdS_5\times S^5$ boundary at $\partial {\cal A}$:
\begin{equation}\label{eq:EE1}
S_{\cal A}= {A_{\mathrm{min}} \over 4 G_N^{(10)}}
\end{equation}
Since we are dealing with static states of our CFT, this surface lies on a constant time slice. If this surface is not unique, we choose the one whose area is minimal among all such surfaces homologous to ${\cal A}$.\footnote{This minimal surface prescription  was recently established on a firm footing by the analysis of \cite{Lewkowycz:2013nqa}.} 

In the following section we derive the minimal surface ${\cal M}$ for a general bubbling solution and show that its restriction to the boundary, which is a theory on $AdS_3\times S^1$, maps to a two-sphere in the Weyl-related  $\mathbb{R}^4$. We then evaluate its regulated area.

\subsection{Minimal surface geometry}

A bubbling geometry is a $AdS_3\times S^3\times U(1)$ fibration over $X$. We consider a surface ${\cal M}$ at constant $t$ that fills the $S^3$ and has profile $z=z(l,\chi,y,x_1,x_2)$, where $z$ is the $AdS_3$ radial coordinate defined in section~\ref{sec3}.  The induced metric on ${\cal M}$ is
\begin{align}\label{eq:inducedmetric}
h_{\alpha\beta}\, dx^\alpha\, dx^\beta &=  y \sqrt{\frac{2 f + 1}{2 f - 1}}\, \frac{1}{z^2} \left[ dl^2 + \left( \frac{\partial z}{ \partial l}\, dl+\frac{\partial z}{ \partial \chi}\, d\chi+\frac{\partial z}{ \partial y}\, dy+\frac{\partial z}{ \partial x_1}\, dx_1+\frac{\partial z}{ \partial x_2}\, dx_2  \right)^2\right] \nonumber\\  
 &\phantom{=\ } + y \sqrt{ \frac{2 f - 1}{2 f + 1} }\, ds^2_{S^3} + \frac{2 y}{\sqrt{4f^2-1}} \left[ d\chi^2 + 2 V_I\, dx^I\, d\chi +\left(V_I\, dx^I\right)^2 \right] \nonumber\\  
 &\phantom{=\ } +  \frac{\sqrt{4f^2-1}}{2 y}\, \left( dy^2 + dx_1^2 +  dx_2^2 \right)
\end{align}
where $\alpha,\beta$ run over all coordinates except $t$ and $z$. The area functional becomes 
\begin{align}\label{eq:areaintegral}
A({\cal M}) &=  \textrm{Vol}\left(S^3\right) 
\int dl\, d\chi\, dy\, dx_1\, dx_2 \, \frac{\left(f-\tfrac{1}{2}\right)y}{z}\,   \left\{1+ \left( \frac{\partial z}{ \partial l}\right)^2 +\frac{y^2}{\left(f-\tfrac{1}{2}\right) z^2} \left[ 
 \left( \frac{\partial z}{ \partial y}\right)^2 \right. \right. \nonumber\\
 &\phantom{=\ } 
 \left. \left.  + \left(\frac{\partial z}{\partial x_1} -V_1\,  \frac{\partial z}{  \partial \chi}\right)^2 +\left(\frac{\partial z}{\partial x_2} -V_2\,  \frac{\partial z}{  \partial \chi}\right)^2  +\frac{\left(f+\tfrac{1}{2}\right) \left(f-\tfrac{1}{2}\right)}{y^2} 
 \left( \frac{\partial z}{ \partial \chi}\right)^2
 \right]    \right\}^{\frac{1}{2}}
\end{align}
The equation of motion that follows from this functional is very complicated, but can be solved by
\begin{equation}\label{eq:surface}
z(l,\chi,y,x_1,x_2)^2 + l^2 = R^2
\end{equation}
This semicircle is a co-dimension two minimal surface in $AdS_3$. Following  \cite{Jensen:2013lxa,Estes:2014hka}  one can show that within this ansatz this is in fact the surface of  minimal area.

The surface \eqref{eq:surface} is independent of the $AdS_5$ radial coordinate. Thus, the boundary $\partial {\cal A}$ of the entangling region on $AdS_3\times S^1$ satisfies the same formula. To understand this better, let us consider two coordinate charts on $\mathbb{R}^4$:
\begin{align}
\label{eq:Weylrescaling}
ds^2_{\mathbb{R}^4}&= z^2 \left( \frac{dz^2+dt^2+dl^2}{z^2} + d\psi^2\right) = dt^2 + dx^2 +x^2 \left(d\vartheta^2 +\sin^2\vartheta\,  d\psi^2 \right)
\end{align}
The map between these two charts is given by
\begin{align}
z&=x \, \sin\vartheta,\quad\quad l = x\, \cos\vartheta
\end{align}
Thus, our entangling surface $\partial {\cal A}$ on the space $AdS_3\times S^1$ can be written as a two-sphere of radius $R$ on $\mathbb{R}^4$ (given by $x=R$) upon Weyl rescaling.

\subsection{Evaluating the area integral}

The minimal area can be written as follows:
\begin{equation}\label{eq:minimalarea}
A_{\mathrm{min}} =  \textrm{Vol}\left(S^3\right) \textrm{Vol}\left(S^1\right) \int dl\, \frac{R}{R^2-l^2}\,  I 
\end{equation}
where we have defined
\begin{equation}\label{eq:I}
I \equiv \int_X  dy\, dx_1\, dx_2 \left(f-\tfrac{1}{2}\right) y 
\end{equation}
with the function $f$ given in \eqref{fforma}. The area integral, and hence the entanglement entropy, diverges. This is expected due to the infinite number of degrees of freedom localized near the entangling surface and is present even in the vacuum. However, the intersection between the entangling surface and the surface operator leads to an additional divergence. Our goal is to extract the change in entanglement entropy in the presence of the surface operator, which requires a  careful treatment of these divergences.

We introduce two independent cut-offs, which we now argue is consistent with our field theory living on $AdS_3\times S^1$.  Firstly, the integral over $X$ diverges due to the infinite volume of $AdS_5$.  We regulate this with our Fefferman-Graham cut-off $\varepsilon$, which is a UV cut-off on $AdS_3\times S^1$.  Secondly, after using \eqref{eq:surface} to rewrite the $l$ integral as an integral over $z$, we find a divergence at  $z=0$.  This is the location of the surface operator and is at infinite proper distance from other points in the $AdS_3$. We therefore interpret this as an IR cut-off and regulate at $z=\eta$. 

It is instructive to focus first on the case with no  surface operator present in order to exhibit the divergence structure of  these integrals most clearly.  We begin by changing coordinates via \eqref{eq:changecoordinates}.  Defining $\alpha\equiv \psi+\phi$ and using the vacuum formula \eqref{eq:vacuumfandV} for $f$ we find
\begin{equation}\label{eq:vacuumI}
I^{(0)} =  \int_0^{2\pi} d\alpha \int_0^{\pi/2} d\theta\, \cos^3\theta\, \sin\theta \int_0^{\rho^{(0)}_c} d\rho\,  \rho 
\end{equation}
We  denote by $\rho^{(0)}_c$ the Fefferman-Graham cut-off function \eqref{eq:cut-off} evaluated on the vacuum moments \eqref{eq:vacmoments}.  In this special case it truncates to just two terms and is in fact independent of  the angular coordinates: $\rho_c^{(0)}=1/\varepsilon-\varepsilon/4$. Reinstating the overall factor of $L^8$, the full result for the integral over $X$ is then
\begin{equation}\label{eq:vacuumIresult}
I^{(0)} = L^8 \left( \frac{\pi }{4\varepsilon^2} - \frac{\pi }{8} + \frac{\pi \varepsilon^2}{64} \right)
\end{equation}
Next we handle the integral over $l$.  Recall that the minimal surface formula \eqref{eq:surface} describes a semicircle for which $z\in[0,R]$ and $l\in[-R,R]$.  The $l$  integral diverges at both limits; rewriting via \eqref{eq:surface}  as an integral over $z$, we regulate with a cut-off at $z=\eta$:
\begin{align}
\int_{-\sqrt{R^2-\eta^2}}^{\sqrt{R^2-\eta^2}} dl\, \frac{R}{R^2-l^2}&=2 \int_0^{\sqrt{R^2-\eta^2}} dl\, \frac{R}{R^2-l^2} = 2  \int_{\eta}^R dz\, \frac{R}{z \sqrt{R^2-z^2}} \nonumber\\
 &= 2 \log\left(\frac{R + \sqrt{R^2 - \eta^2}}{\eta}\right) = 2\log\left(\frac{2R}{\eta}\right) - \frac{\eta^2}{2 R^2} + O\left(\eta^4\right)\label{eq:AdS3cut}
\end{align}
To compute the entanglement entropy \eqref{eq:EE1} we need the following relations between gravity and gauge theory quantities
 \begin{equation}\label{eq:GandLdefinitions}
4 G_N^{(10)} = (2\pi)^7 (4\pi)^{-1}g_s^2 \alpha'^4, \quad L^4 = 4\pi g_s N \alpha'^2
\end{equation}
as well as the volume $ \textrm{Vol}\left(S^3\right) = 2\pi^2$.
Our final result for the divergent terms of the entanglement entropy in the absence of the surface operator is 
 \begin{equation}
S_{\cal A}^{(0)} = N^2 \left[\frac{1}{\varepsilon^2} - \frac{1}{2} + O\left(\varepsilon^2\right) \right] \log \left( \frac{2R}{\eta} \right)
\end{equation}

This result looks very different to that for a spherical entangling surface on $\mathbb{R}^4$  with a single Poincar\'{e}-invariant UV cut-off (see   \cite{Ryu:2006bv}, for example).  The reason is that the $AdS_5$ boundary in the slicing \eqref{eq:AdSmetric} can be reached in two ways: $z\to 0$ at fixed $\rho$  (the location of the surface defect) or $\rho\to\infty$ at fixed $z$ (some point away from the defect).  We therefore need two cut-offs in this chart.\footnote{This situation is also familiar from the $S^1\times H^{d-1}$ slicing of $AdS_{d+1}$ --- see figure~1 of \cite{Karch:2014ufa}, for example.}  For a field theory  on $AdS_3\times S^1$, the cut-off   $\eta$ can be viewed as an IR cut-off that regulates the infinite volume of $AdS_3$.  As we will discuss in some detail in section \ref{sec6}, from the point of view, of the surface defect $\eta$ should be viewed as a UV cut-off.

Now let us evaluate the area integral in the  presence of a surface operator.  Our result \eqref{eq:AdS3cut} for the integral over $l$ is unchanged. Whilst it is possible to evaluate the integral for $I$ given in \eqref{eq:I} for a general bubbling geometry after changing coordinates via \eqref{eq:changecoordinates},  the result is extremely lengthy and cumbersome to deal with.  We found the following  approach to be much simpler.

For a general bubbling geometry, the integral \eqref{eq:I} is actually a sum of integrals:
\begin{equation}\label{Iintermoff}
I=\sum_{i=1}^M I_i \quad \textrm{with} \quad I_i\equiv \int_X dy\, dx_1\, dx_2\, y f_i 
\end{equation}
where the $f_i$ are given in \eqref{eq:fiformula}. We can perform a change of variables for each value of $i$ separately
\begin{equation}\label{eq:barredvariables}
x_1 = y_i\, \bar x_1 + x_{i1}, \quad x_2 = y_i\, \bar x_2 + x_{i2}, \quad y=y_i\, \bar y
\end{equation}
after which the $I_i$ integral becomes
\begin{equation}
I_i = y_i^4 \int_{\bar{X}} d\bar y\, d\bar x_1\, d\bar x_2 \, \bar y\, f_i \quad \textrm{with} \quad  f_i = -\frac{1}{2} + \frac{\bar y^2+\bar x_1^2+\bar x_2^2+1}{2\sqrt{\left(\bar y^2+\bar x_1^2+\bar x_2^2+1\right)^2-4\bar y^2}}
\end{equation}
Now $f_i$ takes the same form as for the vacuum configuration.  With a further change of variables the integral can be brought into the same form as \eqref{eq:vacuumI}:
\begin{gather}
I_i= y_i^4 \int  d\bar \alpha\, d\bar \theta\, d\bar \rho \, \bar \rho\, \cos^3 \bar \theta\, \sin \bar \theta\label{eq:Iiinbarredcoords} \\
\bar y = \sqrt{\bar \rho^2+1} \, \cos\bar \theta, \quad \bar x_1= \bar \rho\,  \sin\bar \theta\, \cos\bar \alpha, \quad  \bar x_2= \bar \rho\, \sin\bar \theta\, \sin\bar \alpha \label{eq:barredvariables2}
\end{gather}
All that remains is to impose the correct cut-off in the new variables $\bar{\rho}_c(\varepsilon,\bar\theta,\bar\alpha)$ and then sum up the results for each $I_i$.

As a side remark, it  is interesting that we can express  the general integral in the same form as the vacuum. This is because the function $f$ for the general solution is constructed by superimposing terms that each have the same form as the vacuum solution.  This simple behavior is special to this system and we do not expect such a simplification to be possible generically.

In order to find $\bar{\rho}_c(\varepsilon,\bar\theta,\bar\alpha)$, our strategy is first to express the unbarred variables  $\{  \rho, \theta,\alpha\}$ in terms of the barred variables $\{ \bar \rho,\bar \theta,\bar \alpha\}$ then to write the FG coordinate $u$ as an asymptotic series in large ${\bar \rho}$. Solving this relation asymptotically for $\bar\rho$ and setting $u=\varepsilon$ we obtain the following  cut-off function:
\begin{align}
\bar{\rho}_c(\varepsilon,\bar\theta,\bar\alpha) &= \frac{1}{y_i\, \varepsilon} - \frac{r_i \cos\left(\bar\alpha+\beta_i\right) \sin\bar\theta}{y_i} + \frac{1}{8 y_i} \left[ -1-4r_i^2-2y_i^2- 2(y_i^2-1)\cos2\bar\theta\right. \nonumber\\
& \phantom{=\ } - 2m_{220}-2m_{202}+m_{400} + \sin^2\bar\theta \left(3+2r_i^2+2r_i^2 \cos\left(2\bar\alpha+2\beta_i\right) \right. \nonumber\\
& \phantom{=\ } \left.\left.+6m_{220}+6m_{202}-3m_{400}+12m_{211} \sin 2\bar\alpha +6 \left(m_{220}-m_{202}\right) \cos 2\bar\alpha\right) \right] \varepsilon \nonumber\\
& \phantom{=\ } + O \left( \varepsilon^2\right) 
\end{align}
where we have defined $x_{i1}= r_i  \, \cos\beta_i$ and $x_{i2}= r_i\, \sin\beta_i$. The details on the derivation of the cut-off function $\bar{\rho}_c(\varepsilon,\bar\theta,\bar\alpha)$ are presented in appendix \ref{app:EE}.    Since the coordinate change \eqref{eq:barredvariables} is simply  a rescaling followed by a translation, we deduce the following ranges for the integration variables in the $I_i$ integral \eqref{eq:Iiinbarredcoords}:
\begin{equation}
0 \leq \bar\rho < \bar{\rho}_c(\varepsilon,\bar\theta,\bar\alpha), \quad \bar\theta\in [0,\pi/2], \quad \bar\alpha\in [0, 2\pi]
\end{equation}

We are now ready to evaluate $I_i$. We perform the $\bar \rho$ integral first due to its variable limit.  It turns out that  the moments drop out in the integration over the angular coordinates.  However, they do appear in the final result for $I$ once we sum over $i$: 
\begin{equation}\label{eq:Iresult}
I = \sum_{i=1}^M I_i = \frac{\pi L^8}{4\varepsilon^2}+\frac{\pi L^8}{24} \, \left[1 - 4\left(m_{220}+m_{202}+m_{400}\right)\right]+O\left(\varepsilon\right) 
\end{equation}
where we restored the overall factor of $L^8$. As a leading  order check we do indeed recover the vacuum result \eqref{eq:vacuumIresult} when evaluated on the vacuum moments \eqref{eq:vacmoments}.  The holographic entanglement entropy   in the presence of a surface operator \eqref{eq:EE1} is evaluated using the minimal area via \eqref{eq:minimalarea} in terms of the two regulated integrals  \eqref{eq:AdS3cut} and \eqref{eq:Iresult}. At the end, gravity expressions  are translated to gauge theory ones using \eqref{eq:GandLdefinitions}.  Putting all this together, the  result is 
 \begin{equation}\label{eq:EEgeneral}
 S_{\cal A} = N^2 \left[ \frac{1}{\varepsilon^2} +\frac{1-4\left(m_{220}+m_{202}+m_{400}\right)}{6} +O\left(\varepsilon\right) \right] \log\left(\frac{2R}{\eta}\right)
 \end{equation}
Subtracting the vacuum contribution from \eqref{eq:EEgeneral} and taking $\varepsilon\to 0$ we arrive at our final result for the change in entanglement entropy due to the presence of a  surface operator:
  \begin{equation}\label{eq:EEfinal}
 \Delta S_{\cal A} =  \frac{2N^2}{3} \left(1-m_{220}-m_{202}-m_{400}\right) \log\left(\frac{2R}{\eta}\right)
 \end{equation}

\subsection{A 2D CFT  interpretation}
\label{sec:2Dint}

Let us make a few comments on the form of the result \eqref{eq:EEfinal} for the change in the entanglement entropy.  Note immediately that it diverges as $\eta\to 0$.  This additional divergence was anticipated  due to the intersection between the entangling surface and the surface defect.  The intersection occurs at two points separated by an interval, so it seems natural  for the  divergence to be logarithmic: our result takes the same form as the  entanglement entropy across an interval in the vacuum of a generic two dimensional  CFT \cite{Holzhey:1994we,Calabrese:2004eu}.

 Note that the field theory description of the surface operators  in section~\ref{sec2} did not require any additional 2D degrees of freedom localized at the surface defect.  However, in the original paper~\cite{Gukov:2006jk} an alternative construction of the surface defects  by coupling a nonlinear sigma model on $\Sigma$ to the SYM fields was described.  Such a sigma model could describe the 2D CFT we are looking for in the infrared.  This construction is based on an  intersecting D3-D3' brane system that was  first discussed in \cite{Constable:2002xt}. Alternatively the defect can be realized  by  a probe D3-brane in $AdS_5\times S^5$ with an $AdS_3\times S^1$ worldvolume.  Following Karch and Randall~\cite{Karch:2000gx} and letting holography `act twice' makes it likely that a 2D CFT is described  by the modes on the probe brane.

Consequently it seems possible that the coefficient of the logarithmic divergence in the subtracted entanglement entropy to be equal to (one third of) the central charge of this CFT~\cite{Holzhey:1994we}.
We now provide evidence realizing this expectation.  Recall that our metric~\eqref{metans} takes the form
\begin{equation}
ds^2 = L^2 \left(e^{2W}\, ds^2_{AdS_3} + ds^2_{Z} \right)
\end{equation}
We define an effective central charge via  the Brown-Henneaux fomula \cite{Brown:1986nw}:
\begin{equation}\label{eq:BrownHenneaux}
c_{\textrm{eff}} = \frac{3 L}{2 G_N^{(3)}}
\end{equation}
where $G_N^{(3)}$ is the three-dimensional Newton's constant of the theory obtained by reducing on the remaining directions in $Z$.  To compute $G_N^{(3)}$ we must take into account the non-trivial warp factor in front of $ds^2_{AdS_3}$:
\begin{equation}
 \frac{1}{16\pi G_N^{(3)}} = \frac{1}{16\pi G_N^{(10)}}\, \Delta\left( \int_{Z} d^7x \, \sqrt{g_Z}\, e^{W}\right)
\end{equation}
where in order to isolate the contribution from the surface operator we should subtract off the vacuum answer. Substituting the metric \eqref{metans} and reinstating the correct powers of $L$, our result for the effective central charge via \eqref{eq:BrownHenneaux} is given by
\begin{equation}\label{eq:ceff}
c_{\textrm{eff}} = \frac{3}{2 G_N^{(10)}}\, \textrm{Vol}\left(S^3\right) \textrm{Vol}\left(S^1\right) \Delta I
\end{equation}
where $I$ is the integral  \eqref{eq:I} appearing in the entanglement entropy.  From the minimal area prescription \eqref{eq:EE1} and integral \eqref{eq:minimalarea} we deduce that
\begin{equation}
\Delta S_{\cal A} =  \frac{c_{\textrm{eff}}}{3}\, \log\left(\frac{2R}{\eta}\right)
\end{equation}
which is indeed the entanglement entropy across an interval of length $2R$.  Note that from the point of view of the two dimensional CFT the cut-off $\eta$ is a UV cut-off.

Using  \eqref{eq:ceff} the central charge $c_{\textrm{eff}} $  can be expressed in terms of the moments 
\begin{equation}
c_{\textrm{eff}}= 2N^2\big( 1- m_{400}-m_{220}-m_{202}\big)
\end{equation}
which shows that it 
scales like $N^2$.  This is to be contrasted with the sigma model or probe brane construction mentioned above where one would expect that central charge to scale like $N^0$ or $N^1$, respectively. This result makes sense since the holographic supergravity solution is described by a fully back-reacted geometry in which the number of probe branes scales like $N$, leading to a number of localized degrees of freedom of order $N^2$.
It is also instructive to use the map \eqref{fieldTheoryToSugra} to express $c_{\textrm{eff}}$ in terms of field theory quantities:
\begin{equation}
c_{\textrm{eff}}= 2\left(N^2-\sum_{i=1}^M N_i^2\right) - {8\pi^2 \over g_{YM}^2} \sum_{i=1}^M N_i\left(\beta_i^2+\gamma_i^2\right)
\end{equation}
It is intriguing that the first term agrees with the central charge for the sigma model for an ${\cal N}=(4,4)$ two dimensional quiver gauge theory which is related to a pure monodromy defect (where the $\beta_i$ and $\gamma_i$ vanish), discussed in \cite{Gadde:2013ftv}. It would be very interesting to explore whether the discussion of 
 \cite{Gadde:2013ftv} can be generalized for nonvanishing $\beta_i$ and $\gamma_i$.\footnote{We are grateful to Bruno Le Floch for pointing out  reference  \cite{Gadde:2013ftv} and useful discussions on the possible relation to our results.}

%%%%%%%%%%%%%%%%%%%%%%%%%%%%%%%%
\section{Holographic expectation values}
\setcounter{equation}{0}
\label{sec5}
%%%%%%%%%%%%%%%%%%%%%%%%%%%%%%%%
This section is devoted to holographic expectation values of different observables. Specifically, we calculate the expectation value of the surface defect $\cO_\Sigma$ at strong coupling and large $N$. Our result \eqref{expectationValueResult} is new and is expressed in terms of the moments we introduced in \eqref{momentsDefinition}. We also quote the result of \cite{Drukker:2008wr} for the holographic one-point function of the stress tensor in the presence of $\cO_\Sigma$ (\ref{stressTensorForm2}, \ref{hResult}). In section~\ref{sec6} we will make use of these two expectation values in an attempt to relate them to the entanglement entropy  computed in section~\ref{sec4}.

\subsection{$\langle \cO_\Sigma \rangle$ calculation}

A holographic calculation for the expectation value of the surface operator relies on evaluating the on-shell ten-dimensional type IIB supergravity action  on the bubbling supergravity solution presented in section \ref{sec3}. The obstacle here is well-known: it is difficult to reconcile Poincar\'e invariance of the action with  the self-duality condition of the five-form $F_5$. Different approaches to this problem have been introduced in the literature: Covariant Lagrangians were constructed with the introduction of an infinite number of auxiliary fields \cite{McClain:1990sx,Wotzasek:1990zr,Martin:1994np,Devecchi:1996cp,Bengtsson:1996fm,Berkovits:1996nq,Berkovits:1996rt}, a single auxiliary field in a non-polynomial way \cite{Pasti:1996vs,DallAgata:1997ju,DallAgata:1998va,DallAgata:1998wh} and most recently a construction with a  free  auxiliary four-form  field \cite{Sen:2015nph}. Formalisms with non-manifest Lorentz symmetry were also considered \cite{Henneaux:1988gg,Schwarz:1993vs,Belov:2006xj}.  The solutions  presented in section \ref{sec3} follow from the  standard IIB action where the  the self-duality constraint \eqref{totalAction} has to be imposed  by hand and not derived from varying the action.

In the holographic approach, the expectation value of the surface operator is given by the on-shell action $I:$
 \begin{equation}\label{expectationValueOnShell}
 \langle \mathcal{O}_\Sigma \rangle = \exp\left[-\left(I-I_{(0)}\right)\right]
 \end{equation}
where we subtract off the vacuum contribution $I_{(0)}$. The total action is a sum of a bulk term and the Gibbons-Hawking  term:
\begin{align}
I &= I_{\textrm{bulk}}+I_{\textrm{GH}}\label{totalAction}\\
I_{\textrm{bulk}} &=\frac{1}{2\kappa^2} \left[ \int d^{10}x \sqrt{-g} \left( R -\frac{1}{2} \frac{\partial_M \tau \; \partial^M \bar{\tau}}{(\Im \tau)^2 }\right) \right.\nonumber \\
&\phantom{=\ }\left.-\int \left(  \frac{1}{2} M_{ab} H_{3}^a\wedge \star H_{3}^b +4 F_5\wedge \star F_5 +\epsilon_{ab}C_4\wedge H_{3}^a\wedge H_{3}^b  \right)\right] \\
I_{\textrm{GH}} &= \frac{1}{\kappa^2} \int d^9 x\, \sqrt{-\gamma} \, K \label{GHterm}
\end{align}
In our case the complex scalar  $\tau$ field is constant  and  the three-forms $H_3^a$ vanish.   The trace of the equation of motion for the metric implies $R=0$ and thus the bulk term reduces to
\begin{equation}\label{bulkTerm}
I_{\textrm{bulk}}=-\frac{2}{\kappa^2} \int  F_5\wedge \star F_5 
\end{equation}

To evaluate the bulk term we have to deal with the self-duality of $F_5$ which when imposed makes (\ref{bulkTerm}) vanish.
In the following we employ a pragmatic  method proposed in \cite{Giddings:2001yu,DeWolfe:2002nn}. The prescription suggests to replace  $F_5$ by its electric part only and double the relevant term in the action.  The electric part of $F_5$ is the component with a time-like leg.
As argued in \cite{Giddings:2001yu,DeWolfe:2002nn}  this approach is consistent with Kaluza-Klein reduction and T-duality. It would be interesting to use some of the alternative approaches to
deal with the self-dual five-form. This would however imply redoing the derivation of the BPS supergravity solutions in the respective formalism, which is a somewhat daunting task.

Thus, instead of \eqref{bulkTerm} we need to evaluate
\begin{equation}\label{actionElectric}
I_{\textrm{bulk}}=-\frac{4}{\kappa^2} \int  F_5^{\textrm{el.}}\wedge \star F_5^{\textrm{el.}}
\end{equation}
As the electric part $ F_5^{\textrm{el.}}$ is not self-dual , the integrand of \eqref{actionElectric} does not vanish in general. In particular,  since the time coordinate lies in the $AdS_3$, the electric part  of $F_5$ in  \eqref{F5Solution} consists of the terms that have legs on $AdS_3$. It follows from the self-duality of $F_5$  that the Hodge dual of  $F_5^{\textrm{el.}}$  is the magnetic piece of $F_5$, which  has legs in $S^3$. Consequently we get
 \begin{align}
F_5^{\textrm{el.}}&=- {1\over 4} \Big( d\Big[ y^2{2f+1\over 2f-1}(d\chi +V) \big]- y^3 \star_3 d\Big[ {f+1/2\over y^2}\Big]\Big) \wedge \omega_{AdS_3} \\
\star F_5^{\textrm{el.}}&=- {1\over 4} \Big( d\Big[ y^2{2f-1\over 2f+1}(d\chi +V) \big]- y^3 \star_3 d\Big[ {f-1/2\over y^2}\Big]\Big) \wedge \omega_{S_3}
 \end{align}
 Using the equation \eqref{vdeterm} for the one-form $V$ we can write the integrand in \eqref{actionElectric} as
 \begin{align}
 F_5^{\textrm{el.}}\wedge\star F_5^{\textrm{el.}} =& -\frac{y f}{2(1-4f^2)^2} \Big[ 1- 8f^2+16f^4+\frac{2y}{f}(1-4f^2) \partial_y f \\
 &+4y^2\left((\partial_1 f)^2+(\partial_2 f)^2+(\partial_y f)^2\right)\Big] \omega_{AdS_3}\wedge \omega_{S_3}\wedge d\chi \wedge dx_1\wedge dx_2\wedge dy
 \end{align}
 which can be rewritten in the following way:
  \begin{align}\label{actionElectricIntegrand}
 F_5^{\textrm{el.}}\wedge \star F_5^{\textrm{el.}} =&\left( -\frac{1}{2} y f + \partial_I u_I+ \frac{y^3}{4(1-4f^2)} \left[\partial_1^2 f+\partial_2^2 f+y \partial_y \left( \frac{\partial_y f}{y}\right) \right] \right) \nonumber\\
 &\times \omega_{AdS_3}\wedge \omega_{S_3}\wedge d\chi \wedge dx_1\wedge dx_2\wedge dy
 \end{align}
 where $I$ labels coordinates which run over the base space $X$, $I=\{x_1,x_2,y \}$ and 
 \begin{align}\label{totalDerivativeVector}
 u_I\equiv -\frac{y^3}{4(1-4f^2)}\, \partial_I f
 \end{align}
Using the equation \eqref{fdiffeq} for $f$, we can eliminate the final term in \eqref{actionElectricIntegrand} since its denominator diverges. This is because $f$ diverges at the location of the sources   $y_i, \vec{x}_i.$ Thus, the expression for the integrand is given by
 \begin{equation}\label{bulkElectricSimplified}
  F_5^{\textrm{el.}}\wedge \star F_5^{\textrm{el.}} =\left( -\frac{1}{2} y f + \partial_I u_I(x_1,x_2,y)  \right)\omega_{AdS_3}\wedge \omega_{S_3}\wedge d\chi \wedge dx_1\wedge dx_2\wedge dy
 \end{equation}
The first term appearing in \eqref{bulkElectricSimplified} includes the holographic entanglement entropy integral \eqref{eq:I}. The last term is a total derivative that can be integrated by applying Stoke's theorem. For the convenience of the reader and completeness we present the evaluation of the integrals for the bulk term in the appendix \ref{appastBulk}. The result found in \eqref{bulkActionResult} is as follows:\footnote{$\cF$ is identical to the expression  $128\Delta\Phi_{2,k}\Delta\Phi_{2,-k}$ with $k=-2,0,2$ appearing in \cite{Drukker:2008wr}. $\Delta\Phi_{2,k}$  are the asymptotic coefficients in a spherical harmonic expansion. Details on this expansion and the relation of $\Delta\Phi_{2,k}$ to our moments can be found in appendix \ref{appstress}.}
\begin{equation}\label{bulkActionResult2}
I_\textrm{bulk}= \frac{\pi}{2 \kappa^2} \,\textrm{Vol}\left(AdS_3\right) \textrm{Vol}\left(S^3\right) \textrm{Vol}\left(S^1\right) \left[ \frac{1}{\varepsilon^4}+\frac{1}{\varepsilon^2}+ \frac{3}{8} -  m_{400}- \cF  \right]
\end{equation}
where $\varepsilon$ is the FG cut-off appearing in \eqref{eq:cut-off}. The final term in the finite piece takes the following form in terms of the moments:
\begin{align}
\cF &\equiv \frac{3}{32} \left[ 1 +4 m_{220}  +4 m_{202}-2m_{400} +10\left(m_{220}^2 +m_{202}^2\right)\right. \nonumber\\
  &\phantom{= \frac{3}{32} \left[ \right.} \left. +24 m_{211}^2-4\left(m_{220}+m_{202}\right)m_{400}+m_{400}^2 -4 m_{220} m_{202} \right] \label{eq:Finmoments}
\end{align}

The computation of the Gibbons-Hawking term is performed in the appendix~\ref{appastGH}. The outcome \eqref{eq:GHresult} is given by
 \begin{align}\label{eq:GHresultQuote}
 I_{\textrm{GH}} = \frac{\pi}{2\kappa^2}\,\textrm{Vol}\left(AdS_3\right) \textrm{Vol}\left(S^3\right) \textrm{Vol}\left(S^1\right) \left(\frac{4}{\varepsilon^4} +  \frac{1}{\varepsilon^2} \right)
\end{align}
We note that the Gibbons-Hawking term does not depend on the moments and is hence independent of the details of the bubbling solution. It is notable that  in the analogous calculation of the expectation value for the Wilson surface operator in six-dimensional $(2,0)$ theories \cite{Gentle:2015jma} the Gibbons-Hawking term is also independent of the moments.

\subsection{Result and comments}

Now we are ready to put all the pieces together to build the total on-shell action \eqref{totalAction}. Our result is
\begin{align}
I = \frac{\pi}{2\kappa^2}\,\textrm{Vol}\left(AdS_3\right) \textrm{Vol}\left(S^3\right) \textrm{Vol}\left(S^1\right) \left[\frac{5}{\varepsilon^4} +  \frac{2}{\varepsilon^2} + \frac{3}{8} -  m_{400}- \cF \right]
\end{align} 

Subtracting the vacuum contribution (which has $m_{400}=1$), reinstating the overall factor of $L^8$ and converting to field theory quantities using \eqref{eq:GandLdefinitions} along with $\kappa^2 = 8 \pi G_N^{(10)}$, we arrive at our final result for the expectation value:
\begin{align}\label{expectationValueResult}
\log\, \langle \mathcal{O}_\Sigma \rangle = \frac{N^2}{(2\pi)^2}\left(m_{400}-1+\cF \right) \textrm{Vol}\left(AdS_3\right) \textrm{Vol}\left(S^1\right)
 \end{align}
 
 We should compare our  holographic result for the expectation value with the semi-classical field theory calculation given in \cite{Gomis:2007fi}.  There, the SYM action was evaluated on $AdS_3\times S^1$ with the surface defect boundary conditions \eqref{aadsf} and \eqref{phadsf} imposed and it was found that  $\log\, \langle \mathcal{O}_\Sigma \rangle =0$. A field theory interpretation of the holographic result \eqref{expectationValueResult} in the weak coupling limit is not direct. This is since our result is evaluated using holography and it is valid at strong coupling and large $N$. Even though the surface operator preserves supersymmetry it is not clear that the holographic results can be trusted at weak coupling.  For completeness, however, we make use of the identifications  \eqref{fieldTheoryToSugra} and \eqref{eq:GandLdefinitions}  to express the moments appearing in \eqref{expectationValueResult} in terms of field theory quantities:
 \begin{align}
 m_{400}&=\sum\limits_{i=1}^M \frac{N_i^2}{N^2} 
 \end{align}
 and
 \begin{align}
 \frac{\cF}{6144}&= \left[ \frac{1}{2} -\frac{1}{2 N^2} \sum\limits_{i=1}^M N_i^2 +\frac{4 \pi^2}{g_{YM}^2 N^2} \sum\limits_{i=1}^M N_i \left(\beta_i^2+\gamma_i^2\right)\right]^2\nonumber\\
&\phantom{=\ }+ \frac{24 \pi^4}{g_{YM}^4 N^4} \sum\limits_{i=1}^M N_i\left(\beta_i + i \gamma_i \right)^2  \sum\limits_{j=1}^M N_j\left(\beta_j- i \gamma_j \right)^2
 \end{align}
The interpretation of ${\cF}$ in the field theory is not clear at this point. One would expect that this term should be a higher order correction to the semi-classical  calculation of \cite{Gomis:2007fi} and it would be interesting to calculate quantum corrections to surface  defect operators systematically.

\subsection{$\langle T_{\mu\nu}\rangle_\Sigma$}
\label{sec:stresstensor}

Here we present the stress-energy tensor $\langle T_{\mu\nu}\rangle_\Sigma$ result, evaluated in \cite{Drukker:2008wr}, which we use in the next section. Conformal symmetry constrains the stress-energy tensor form in the presence of the surface defect $\cO_\Sigma$ to \eqref{stressTensorForm}:
\begin{align}\label{stressTensorForm2}
\langle T_{\mu\nu}\rangle_\Sigma \;dx^\mu dx^\nu = h_\Sigma \left(ds^2_{AdS_3} - 3 \, d\psi^2\right)
\end{align}
$\< T_{\mu\nu}\>_\Sigma$ is preserved and traceless, in line with the fact that Weyl anomaly vanishes for $AdS_3 \times S^1$.

The exact value of $h_\Sigma$ is calculated in \cite{Drukker:2008wr} following the holographic renormalization method performed in \cite{Skenderis:2007yb}. 
We give the dictionary of the result of \cite{Drukker:2008wr}  in terms of the moments \eqref{momentsDefinition} in appendix \ref{appstress}. The final result for $h_\Sigma$ then takes the following form
\begin{align}\label{hResult}
h_\Sigma = \frac{N^2}{2\pi^2} \left[  \frac{1}{16} - \frac{1}{3}\left( m_{220}+ m_{202} +\frac{1-m_{400}}{2} \right)  \right]
\end{align}

%%%%%%%%%%%%%%%%%%%%%%%%%%%%
\section{Comparing  entanglement entropies}
\setcounter{equation}{0}
\label{sec6}
%%%%%%%%%%%%%%%%%%%%%%%%%%%%

Our main result in this paper is the subtracted entanglement entropy \eqref{eq:EEfinal} calculated in section \ref{sec4}. The geometric setup is easier to visualize in $\mathbb{R}^{4}$ where the spherical entangling surface  is a sphere.  The setup on   $\mathbb{R}^{4}$  is related to $AdS_3\times S^1$ by a diffeomorphism and a Weyl rescaling. We review the various coordinate systems and the geometry of the entangling surface and surface defect in appendix~\ref{appCoordMaps}. 

 In fact, spherical entangling surfaces are special, since the corresponding modular Hamiltonian is (an integral of) a local operator. In \cite{Casini:2011kv}, the authors used this fact to write the entanglement entropy across a spherical entangling surface  of radius $R$ on $\mathbb{R}^{1,d-1}$ as a thermal entropy on the hyperbolic spacetime $\mathbb{R}\times H^{d-1}$. The latter is conformally related to the causal development of the entangling region on the original Minkowski spacetime.

In \cite{Lewkowycz:2013laa} this mapping of entanglement entropy to thermal entropy was applied to the calculation of entanglement entropy in the presence of Wilson loops in ${\mathcal{N}=4}$ SYM theory and ABJM theories. In particular, it was shown that the additional entanglement entropy due to the presence of the Wilson loop can be calculated from the expectation value of the Wilson loop and the one-point function of the stress tensor.  The formula for the additional entanglement entropy due to the presence of a 
Wilson loop is given by\footnote{Note that we use the opposite sign convention for the stress tensor from the one used in \cite{Lewkowycz:2013laa}. Specifically, our convention makes use of the definition $T_{\mu\nu}=\frac{2}{\sqrt g} \frac{\delta S}{\delta g^{\mu \nu}}$.}
\begin{align}\label{eq:LM}
\Delta S &= \log \langle W\rangle - \int\limits_{S^1 \times H^{d-1}}  d^d x\, \sqrt g\, \Delta \langle T_{\tau\tau} \rangle_{W}
\end{align}
where $\Delta \langle T_{\tau\tau} \rangle_{W}$ denotes the subtracted (by the one-point function without the Wilson loop inserted) time component of the stress tensor. The two expectation values in \eqref{eq:LM} are calculated on the hyperbolic space $S^1 \times H^{d-1}$, where the coordinate of the thermal circle $S^1$ is denoted by $\tau\sim\tau+\beta$ with periodicity $\beta=2\pi R$.

The formula \eqref{eq:LM} is valid for arbitary representations of the Wilson surface. If the representation becomes very large, i.e.\ the associated Young tableaux have $N^2$ boxes, the  backreaction on the dual supergravity solution cannot be neglected. This case was examined in~\cite{Gentle:2014lva} by two of the present authors. 
There, the holographic entanglement entropy was calculated using the bubbling supergravity solutions dual to half-BPS Wilson loops  \cite{D'Hoker:2007fq}.  The expectation values reduce by localization  to  matrix model  integrals \cite{Pestun:2007rz}. Once matrix model and supergravity solution data are appropriately identified, following \cite{Okuda:2008px,Gomis:2008qa}, it was found that the holographic entanglement entropy exactly agrees with \eqref{eq:LM}.

We are also studying a setup with a spherical entangling surface in a CFT, so it is interesting to see whether the same formula \eqref{eq:LM} can be applied to our system.  (Of course, the map to a thermal entropy \cite{Casini:2011kv} should still hold because the isometry in $\tau$ is unbroken.) Here, the Wilson loop operator is replaced by a surface defect. To evaluate  \eqref{eq:LM} we have to calculate the  values of $\langle\mathcal{O}_\Sigma\rangle$ and the stress tensor on  $S^1\times H^3$. In section \ref{sec5} we   determined them on $AdS_3\times S^1$, so the first step is map these quantities to the hyperboloid.\footnote{These are the same  Euclidean geometry. However, we wish to map a theory quantized on the time $t$ in $AdS_3$ to a theory quantized on the time $\tau$, so we must perform a non-trivial conformal transformation. Furthermore, the geometric location of the surface defect and the entangling surface is exchanged in the two coordinate systems.} 

Our setup admits a simple description in $\mathbb{R}^4$. The three spaces  are conformally related as follows:
\begin{equation}
ds^2_{AdS_3\times S^1} =z^{-2}\, ds^2_{\mathbb{R}^4} = \Omega^2\, ds^2_{S^1\times H^3}
\end{equation}
where the expressions for the 4D metrics (in the coordinate charts of interest) and the conformal factor $\Omega$ are given in \eqref{eq:4dmetrics} and \eqref{eq:OmegaConformalfactor}:
\begin{equation}
ds^2_{S^1\times H^3} = d\tau^2 + R^2 \left( d\rho^2 +\sinh^2\rho\left( d\vartheta^2+\sin^2\vartheta\, d\psi^2\right)\right)\quad\textrm{and}\quad \Omega^2 = \frac{1}{R^2 \sinh^2 \rho \sin^2\vartheta}
\end{equation}
For convenience of the reader, further details on the coordinate maps and the description of our setup in these charts is given in appendix \ref{appCoordMaps}. 

It was shown in \cite{Graham:1999pm} that even-dimensional surface observables suffer from a conformal anomaly.  In particular, the infinitesimal change in the  expectation value of $\mathcal{O}_\Sigma$ is proportional to a linear combination of integrals of the intrinsic and extrinsic curvatures of the surface, whose precise expression is given in equation~(2.9) of~\cite{Drukker:2008wr}. The coefficients in this combination depend on the surface operator and the theory and are generically non-zero. However, the curvature integrals all vanish in our setup of a planar surface at $\partial AdS_3\subset AdS_3\times S^1$, so we conclude that  $\langle \mathcal{O}_\Sigma\rangle$ is   invariant under this conformal transformation.  (Of course,  the 4D trace anomaly also vanishes on this space, as noted in section~\ref{sec:stresstensor}.)

The one-point function of the stress tensor \eqref{stressTensorForm2} transforms in the usual way under a conformal transformation in four dimensions; for example
\begin{align}
\langle\tilde{T}_{\tau\tau}\rangle_\Sigma  &= \Omega^{2} \left[\left(\frac{\partial t}{\partial \tau}\right)^2 \langle T_{tt}\rangle_\Sigma+\left(\frac{\partial l}{\partial \tau}\right)^2 \langle T_{ll}\rangle_\Sigma+\left(\frac{\partial z}{\partial \tau}\right)^2 \langle T_{zz}\rangle_\Sigma \right]\nonumber \\
 &= \frac{h_\Sigma}{R^4\, \sinh^4\rho\, \sin^4\vartheta}
\end{align}
where we used the coordinate map from $AdS_3\times S^1$ to the hyperboloid in \eqref{eq:4dCoordinateMaps}. The full result is  traceless as expected since the trace anomaly vanishes on $S^1\times H^3$:
\begin{equation}
\langle\tilde{T}_{\mu\nu} \rangle_{\Sigma}\, d\tilde{x}^{\mu}\, d\tilde{x}^{\nu} = \frac{h_\Sigma}{R^4\,  \sinh^4\rho\, \sin^4\vartheta} \left[ d\tau^2 + R^2\left( d\rho^2 +\sinh^2\rho\left(d\vartheta^2 -3 \sin^2\vartheta\, d\psi^2 \right) \right)\right] 
\end{equation}
Note that in even dimensions there is also an inhomogeneous term that generalizes the Schwarzian derivative for the two-dimensional stress tensor. As pointed out in \cite{Casini:2011kv,Hung:2011nu} this term does not depend on   the state of the theory.  Hence  it will drop out of the vacuum subtracted stress tensor component  $\Delta \langle \tilde{T}_{\tau\tau}\rangle_{\Sigma}$ in \eqref{eq:LM}.

For reasons that will become clear later we write the volume factors in the expression of the expectation value, \eqref{expectationValueResult}, in integral form and change variables. The new variables are the coordinates on the hyperboloid, $\{ \tau,\rho,\vartheta,\psi \}$, which have one-to-one map with $AdS_3\times S^1$ coordinates, $\{ t,l,z,\psi \}$. The volume is written as
\begin{align}
\textrm{Vol}\left(AdS_3 \times S^1\right)&=\int\limits_{AdS_3 \times S^1} d^4 x\sqrt{g} = \int\limits_{S^1\times H^3} d^4 \tilde{x}\, \Omega^{-2}\, \sqrt{\tilde{g}} \nonumber\\
&=\frac{ \beta\,  \textrm{Vol}\left(S^1\right)}{R} \int \frac{d\vartheta}{\sin^3\vartheta} \int \frac{d\rho}{\sinh^2\rho} 
\end{align}
where the integration over $\psi$ and the thermal cycle have been performed. We omit the limits of the integrals over $\vartheta$  and $\rho$ to treat them later. Substituting this relation into \eqref{expectationValueResult} we write the expectation value as
 \begin{equation}\label{eq:LMsecond}
\log\, \langle \mathcal{O}_\Sigma \rangle = N^2\left(m_{400}-1+\mathcal{F} \right) \int \frac{d\vartheta}{\sin^3\vartheta} \int \frac{d\rho}{\sinh^2\rho} 
\end{equation}
The third ingredient in \eqref{eq:LM} is (dropping tildes)
\begin{equation}\label{eq:LMthird}
\int\limits_{S^1\times H^3} d^{4}x\, \sqrt{g}\, \Delta\< T_{\tau \tau}\>_\Sigma =  (2\pi)^2\, \Delta h_\Sigma \int \frac{d\vartheta}{\sin^3\vartheta} \int \frac{d\rho}{\sinh^2\rho}
\end{equation}
where $\Delta h_\Sigma$ is the vacuum subtracted value of \eqref{hResult}. 

\begin{figure}[!t]
  \centering
  \includegraphics[width=110mm]{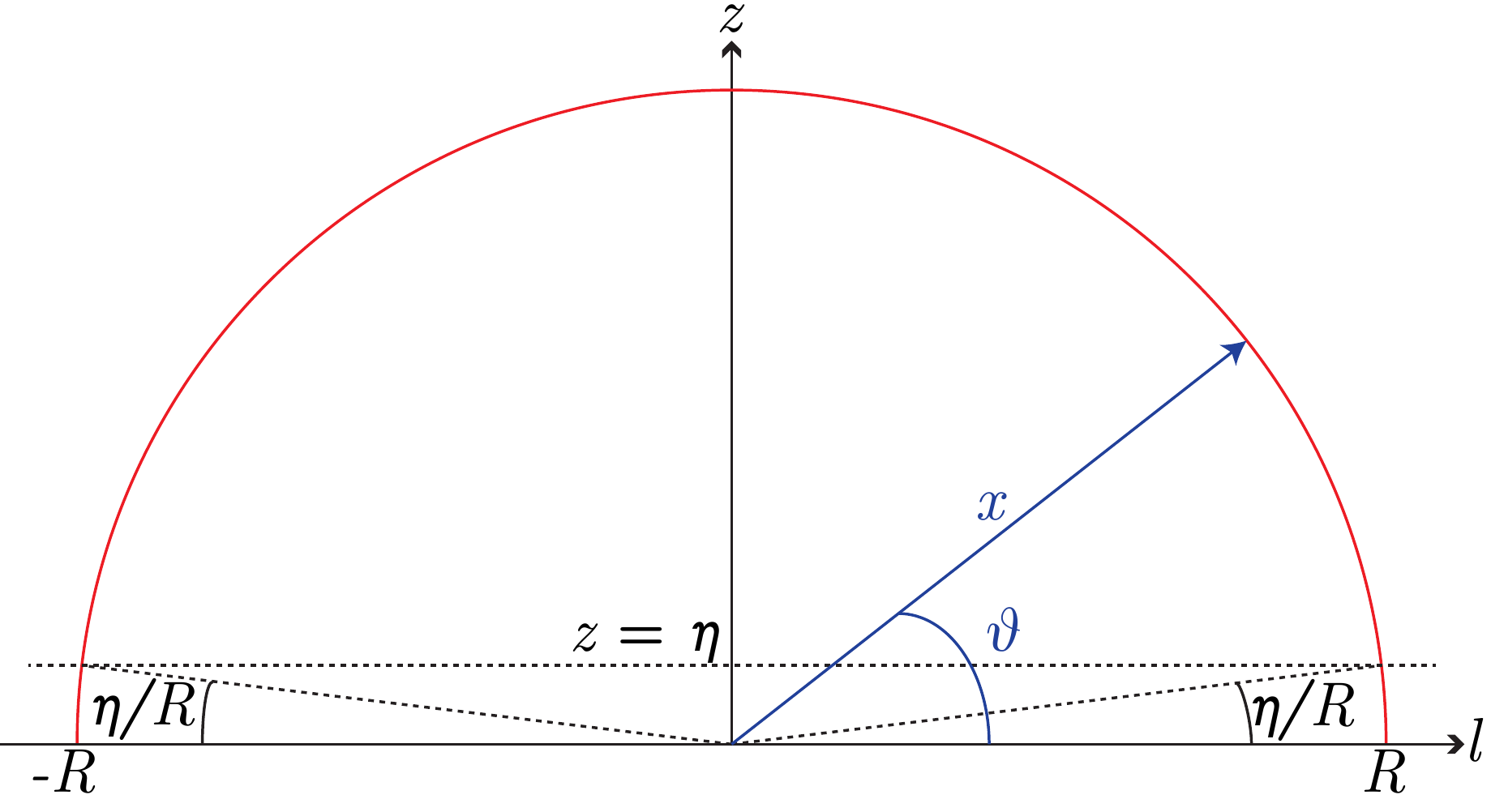}
  \caption{Mapping the $z=\eta$ cut-off to polar coordinates. The red semicircle is the entangling surface while the location of $\cO_\Sigma$ is at $z=0$. A uniform cut-off $z=\eta$ close to the location of $\cO_\Sigma$ is introduced. It is denoted with a dashed horizontal line. This limits integration over $\vartheta$ between $\eta/R$ and $ \pi-\eta/R$.}
 \label{fig:etacutoff}
\end{figure}

We notice that both ingredients (\ref{eq:LMsecond}, \ref{eq:LMthird}) contain the same integrals. The integrals diverge since the domain of integration is $\vartheta\in[0,\pi]$ and $\rho\in[0,\infty)$. To compute them we introduce two independent cut-offs as follows:
 \begin{equation}\label{integralsRhoTheta}
\int_{\eta/R}^{\pi-\eta/R} \frac{d\vartheta}{\sin^3\vartheta} \int_{a}^\infty \frac{d\rho}{\sinh^2\rho}= \left[ \frac{R^2}{\eta^2} +\log\left( \frac{2R}{\eta}\right)-\frac{1}{6}+O\left(\eta^2\right)\right] \left( \frac{1}{a}-1+O(a) \right)
\end{equation}
The cut-off $\eta$ is identified with the homonymous cut-off introduced in the holographic entanglement entropy calculation. The divergence comes from degrees of freedom close to the entangling surface $x=R$. Therefore, for small $z=\eta$ the first map in \eqref{eq:4dCoordinateMaps} sets the cut-off values of $\vartheta$ to $\eta/R$ and $\pi-\eta/R$  (see figure~\ref{fig:etacutoff}). 
Since we are interested in the universal term of \eqref{integralsRhoTheta} where $a$ is absent, no identification for this cut-off is needed.

We are now ready to combine all the ingredients in \eqref{eq:LM} (with the Wilson loop  replaced by the surface defect). The right hand side  is
\begin{align}\label{LMApplied}
\log\, \langle \mathcal{O}_\Sigma \rangle- \int\limits_{S^1\times H^3} d^{d}x\, \sqrt{g}\, \Delta\< T_{\tau \tau}\>_\Sigma  &=  \frac{2N^2}{3}\left(1-m_{220}-m_{202}-m_{400} - \frac{3}{2}\, \mathcal{F} \right) \log\left(\frac{2R}{\eta}\right)\nonumber\\
&=\Delta S_\cA -N^2\, \mathcal{F} \log\left(\frac{2R}{\eta}\right)
\end{align}
We immediately notice that there is a discrepancy compared to \eqref{eq:LM}. The mismatch amounts to the second term in \eqref{LMApplied}, which is proportional to $\mathcal{F}$. The minimal relation \eqref{eq:LM},  derived in \cite{Lewkowycz:2013laa} for Wilson loops, does not work here. This mismatch and possible explanations for it  are discussed further in section \ref{sec7}.

%%%%%%%%%%%%%%%%%%%%%%%%%%%%
\section{Discussion}
\setcounter{equation}{0}
\label{sec7}
%%%%%%%%%%%%%%%%%%%%%%%%%%%%

In this paper we studied two-dimensional planar surface defects in ${{\cal N}=4}$ SYM theory via their dual supergravity bubbling description.  First we computed the entanglement entropy across a ball-shaped region bisected by a surface defect.  In addition we  calculated two other holographic observables: the one-point function of the stress tensor and the expectation value of the surface defect.

We attempted to combine these ingredients as a test of field theory expectations.  After a conformal transformation, our entanglement entropy should be equal to the thermal entropy on a hyperbolic space.  However, as discussed in section~\ref{sec6}, a straightforward generalization of the work of Lewkowycz and Maldacena for Wilson loops  \cite{Lewkowycz:2013laa} did not work in our setup.  We now offer some possible reasons for why this is so.

It could be that our cut-off prescription was too naive.  However, whilst the sub-leading universal term in \eqref{integralsRhoTheta} could receive corrections in a modification of this prescription, the form of the mismatch (characterized by ${\cal F}$ in \eqref{LMApplied}) is very different to that of the other terms, so this cannot be the whole story.

The two new elements in our setup compared to~\cite{Lewkowycz:2013laa} were the conformal anomaly for even-dimensional surface observables and the intersection between the entangling surface and the defect.  
So one possibility is that either element should contribute an extra term to the thermal entropy in addition to those we considered.  The same two elements were present in our previous calculations~\cite{Gentle:2015jma} for a Wilson surface  in the six-dimensional $(2,0)$~theory.  Whilst we do not have a general closed-form expression for the expectation value of the Wilson surface, a case-by-case check yields a similar mismatch. It would be very interesting to pin this down in future work with a direct field theory replica trick calculation.

As discussed for example in \cite{Jensen:2015swa,Billo:2016cpy} there can be an additional contribution to the expectation value of the  stress tensor with a defect present which for the case at hand has  the general form
\begin{equation}
\langle T_{\mu\nu}(x)\rangle= \langle T^{\rm bulk}_{\mu\nu}(x)\rangle+\delta^2(x^\perp) \langle T^{\rm defect}_{\mu\nu}(x)\rangle
\end{equation}
While the bulk contribution to this one point function was given in \ref{sec:stresstensor}, we have not been able to calculate the localized contribution holographically, due to the fact that the standard Fefferman-Graham  coordinates break down near the location of the defect as discussed in \cite{Jensen:2013lxa,Papadimitriou:2004rz}.  We are not aware of any holographic calculation  of the defect contribution to the stress tensor in the literature. It would  however  be very interesting to see whether the   extra localized contribution can  lead to a match of the two ways to calculate the entanglement entropy presented in this paper. Furthermore, on general grounds the localized term is governed by the Weyl anomaly 'living' on the surface  operator and hence such a calculation would also determine this anomaly holographically.

In order to compute the required thermal entropy, one must compute the free energy in the presence of the defect, which involves taking a derivative with respect to the inverse temperature $\beta$.  This can be written as a derivative of the field theory Lagrangian with respect to the metric, as utilized in \cite{Lewkowycz:2013laa}. Whilst the origin of such a term is unclear, it would contribute to the entanglement entropy, so its existence (or lack thereof) should be clear from a replica trick calculation.

Another approach that may provide an understanding of the mismatch is to consider the probe brane approximation, where the defect operator is realized as D3 branes with an $AdS_3\times S^1$ worldvolume inside $AdS_5\times S^5$ and the backreaction is neglected or treated perturbatively.
The expectation value is known for both types of surface operator \cite{Drukker:2008wr, Berenstein:1998ij} and the entanglement entropy can be computed to leading order using the results of \cite{Karch:2014ufa} (see also \cite{Jensen:2013lxa, Chang:2013mca, Kontoudi:2013rla}). It would be interesting to determine whether the mismatch we found persists in this approximation.

The logarithmic divergence in the entanglement entropy in the vacuum of a 2D CFT is universal: it depends only on the central charge. 
It is natural to ask whether  the coefficient in our subtracted result \eqref{eq:EEfinal} is similarly universal.  
We computed an effective central charge holographically in section~\ref{sec:2Dint} and found precise agreement with this coefficient.  Indeed, the same agreement is found for the holographic description of a Wilson surface in the $(2,0)$~theory \cite{Gentle:2015jma}.  It would be very interesting to pursue this connection further and develop a 2D CFT description for both types of surface operator.

\section*{Acknowledgements}

We are delighted to thank Xi Dong, Matthew Headrick, Bruno Le Floch, Edgar Shaghoulian and Christoph Uhlemann for useful discussions. This work was supported in part by National Science Foundation grant PHY-13-13986. The work of MG  was in part supported by a fellowship of the Simons Foundation. MG thanks the Institute for Theoretical Physics, ETH Z\" urich, for hospitality while   this work was in progress.

\newpage

%%%%%%%%%%%%%%%%%%%%%%%%%%%%%%%%%%%%%%%%%%%%
%% Appendix 
%%%%%%%%%%%%%%%%%%%%%%%%%%%%%%%%%%%%%%%%%%%%
\appendix 
\section{Fefferman-Graham coordinates}
\label{app:FGCoordinates}
%%%%%%%%%%%%%%%%%%%%%%

%%%%%%%%%%%%%%%%%%%%%%
This section complements the discussion of the FG mapping procedure in section \ref{sec3}. We describe the gauge choice for the one-form $V$ and give the results of the FG coordinate map.

\subsection{Gauge choice}
\label{app:FGCoordinates:GaugeChoice}
As mentioned in section \ref{sec3} we are interested to choose $\omega$ such that $V_\rho=0$. 
In particular, we first need to expand the function 
\begin{equation}
 \omega = \sum_{n=0}^{\infty} \frac{\omega^{(n)}(\theta,\alpha)}{\rho^n}
\end{equation}
where $\alpha = \psi + \phi$ and demand that $V_\rho = 0$ at each order in the $\rho^{-1}$ expansion. This is a gauge choice that kills all $d\rho\, dY$ cross terms with $Y\in \{\psi,\theta,\phi\}$ in the asymptotic expansion of the metric\footnote{Note that $ds_X^2$ defined in \eqref{basemetric} has no $d\rho\, dY$ cross terms.}.Then we fix $\omega^{(0)}$ by demanding that the $d\theta\, d\psi$ and $d\psi\, d\phi$ cross terms vanish at zeroth order for all $M$. 
Considering the expansion of the one-form \eqref{eq:Vdefinition} at large $\rho$:
\begin{align}
V_{I} = \sum\limits_{n=1}^{\infty}\frac{V_I^{(n)}(\theta,\alpha)}{\rho^n}
\end{align}
The result for $\omega$ is given in terms of ${V_I^{(n)}}$ coefficients in \eqref{omegaResultImplicit}. Substituting the explicit expressions for the coefficients it can be written as
\begin{align}
\omega   &=- \frac{M-1}{2}\, \alpha  + \frac{1}{2\sin\theta\, \rho}\sum_{i=1}^M \left(x_{i2}\cos\alpha-x_{i1}\sin\alpha\right)\nonumber \\
&\phantom{=\ }  -   \frac{1}{4\sin^2\theta\, \rho^2}\sum_{i=1}^M \left[\left(x_{i1}^2-x_{i2}^2\right)\sin 2\alpha-2 x_{i1}x_{i2}\cos 2\alpha\right]\nonumber\\
&\phantom{=\ }  -   \frac{1}{6\sin^3\theta\, \rho^3}\sum_{i=1}^M \left[\left(x_{i1}^3-3x_{i1} x_{i2}^2\right)\sin 3\alpha + \left(x_{i2}^3-3x_{i1}^2 x_{i2}\right)\cos 3\alpha\right] \nonumber\\ 
&\phantom{=\ } +\frac{1}{8\rho^4} \left\{ \frac{1}{\sin^4\theta}\sum_{i=1}^M \left[4  \left(  x_{i1}^3 x_{i2}-x_{i1} x_{i2}^3\right) \cos 4 \alpha- \left(x_{i1}^4 - 
     6 x_{i1}^2 x_{i2}^2 + x_{i2}^4\right) \sin 4 \alpha \right] \right. \nonumber\\
&\phantom{=\ } \left.  + \sin^2\theta\sum_{i=1}^M \left[-8 y_i^2x_{i1} x_{i2} \cos 2 \alpha + 
 4 \left(y_i^2x_{i1}^2 - y_i^2 x_{i2}^2\right) \sin 2 \alpha \right] \right\}+ O\left(\rho^{-5}\right) \label{eq:omega}
\end{align}
This is the gauge choice which eliminates the $V_\rho$ component and brings the metric in a manifestly asymptotically $AdS_5 \times S^5$ form.

\subsection{The coordinate map}
\label{app:FGCoordinates:CoordinateMap}
In this subsection we give the results of the FG mapping. We express them in terms of the expansion coefficients of the functions $F_a$ appearing in \eqref{largeRhoMetric}. The coefficients relevant to our calculation  come from the expansion of $F_\rho$:
\begin{align}
F_\rho = \sum_{n=1}^{\infty} \frac{F_{\rho}^{(n)} (\theta,\alpha)}{\rho^n} 
\end{align}
In what follows we express the relevant coefficients in terms of the moments:
\begin{align}
4 F_{\rho}^{(2)} &=  \left(1-3\cos 2\theta\right) \left[ 1 + 2 \left(m_{220} +m_{202}\right) - m_{400}   \right] \nonumber \\
&\phantom{=\ }  +12\left[ \cos 2\alpha \left(m_{220} - m_{202} \right) +2 m_{211}\sin 2\alpha \right] \sin^2\theta\nonumber\\
F_{\rho}^{(3)} &= 3 \left(\sin\theta-\sin 3\theta  \right)  \left[   \left( m_{212} + m_{230} - m_{410} \right)\cos\alpha + \left( m_{221} + m_{203} - m_{401} \right)\sin\alpha  \right] \nonumber\\
&\phantom{=\ } +4 \sin^3\theta \left[ \left( -3 m_{212} + m_{230} \right)\cos 3\alpha -  \left(-3 m_{221} + m_{203}\right)\sin 3\alpha  \right] \nonumber\\
32 F_{\rho}^{(4)} &= -4 \cos^4\theta +\left(5-12\cos 2\theta +15 \cos 4\theta\right) \left(2m_{202}+2m_{220}-m_{400}\right) \nonumber\\
&\phantom{=\ } -16 \left(1+5\cos 2\theta\right) \sin^2\theta \sin 2\alpha \left[ 3 m_{211} +8 \left( m_{213}+m_{231}\right) -12m_{411} \right]\nonumber\\
&\phantom{=\ } -8 \left(1+5\cos 2\theta\right) \sin^2\theta \cos 2\alpha \left[ 3\left( m_{220}-m_{202}\right)+8 \left(m_{240}-m_{204}\right)+12\left(m_{402}-m_{420}\right)\right]\nonumber\\
&\phantom{=\ } -640 \sin 4\alpha \sin^4\theta \left( m_{213}-m_{231}\right)+24\left(3-4\cos 2\theta +5\cos 4\theta -40\cos 4\alpha\sin^4 \theta \right) m_{222}\nonumber\\
&\phantom{=\ } +4\left( 9-12\cos 2\theta+15\cos 4\theta+40\cos 4\alpha \sin^4 \theta \right) \left( m_{204}+m_{240}\right)\nonumber\\
&\phantom{=\ } - 4 \left(3-4\cos 2\theta+5\cos 4\theta\right) \left[6\left(m_{402}+m_{420}\right) -m_{600}\right]\nonumber\\
&\phantom{=\ }-\Big(12 \sin^2 \theta \left[\cos 2\alpha \left( m_{202}-m_{220}\right) -2\sin 2\alpha \;m_{211}\right]  \nonumber\\
&\phantom{=-\Big(\ } -(1-3\cos 2\theta) \left(2m_{202}+2m_{220}-m_{400}\right)\Big)^2
\end{align}

The FG mapping, as described in section \ref{sec3}, gives the following results for the FG coordinates:
\begin{align}\label{eq:FGMap}
u &= \frac{1}{\rho} \left[ 1  + \frac{F_\rho^{(2)}-1}{4\rho^2} + \frac{F_\rho^{(3)}}{6\rho^3} + \frac{16(F_\rho^{(4)}-F_\rho^{(2)}+1)-(\partial_\theta F_\rho^{(2)})^2- (\partial_\phi F_\rho^{(2)})^2 \csc^2\theta  }{128\rho^4} + O\left(\rho^{-5}\right) \right] 
\nonumber \\
\tilde\psi &= \psi  - \frac{\partial_\psi F_\rho^{(2)}}{16\rho^4} - \frac{\partial_\psi F_\rho^{(3)}}{30\rho^5}+ O\left(\rho^{-6}\right) \nonumber \\
\tilde\theta &= \theta  - \frac{\partial_\theta F_\rho^{(2)}}{8\rho^2} - \frac{\partial_\theta F_\rho^{(3)}}{18\rho^3} + \frac{1}{256\rho^4} \left[ -8 \partial_\theta F_\rho^{(4)} + 3 \partial_\phi F_\rho^{(2)}\, \partial_\theta \partial_\phi F_\rho^{(2)} \csc^2\theta \right. \nonumber \\
&\phantom{=\ }\left.- (\partial_\phi F_\rho^{(2)})^2 \cot\theta \csc^2\theta  +\partial_\theta F_\rho^{(2)} \left( 12-4 F_\rho^{(2)} +16 F_4^{(2)} +3 \partial_\theta^2 F_\rho^{(2)}  \right) \right]+O\left(\rho^{-5}\right) \nonumber \\
\tilde\phi &= \phi  - \frac{\partial_\phi F_\rho^{(2)}}{8\sin^2\theta\, \rho^2} - \frac{\partial_\phi F_\rho^{(3)}}{18\sin^2\theta\, \rho^3} + \frac{1}{256\sin^2\theta\, \rho^4}  \left[ -8\partial_\phi F_\rho^{(4)}+3 \partial_\theta F_\rho^{(2)}\, \partial_\theta\partial_\phi F_\rho^{(2)} \right. \nonumber \\
&\phantom{=\ }\left. + \partial_\phi F_\rho^{(2)} \left(12-4 F_\rho^{(2)} +16 F_5^{(2)} +3 \partial_\phi^2 F_\rho^{(2)} \csc^2\theta - 4 \partial_\theta F_\rho^{(2)} \cot\theta \right) \right]+O\left(\rho^{-5}\right) 
\end{align}

\section{Holographic entanglement entropy}
\setcounter{equation}{0}
\label{app:EE}
%%%%%%%%%%%%%%%%%%%%%%%%%%%%%%%%%%%%%%%%%%%%
In this section we present some details of the holographic entanglement entropy calculation performed in section \ref{sec4}. To compute the integrals \eqref{Iintermoff} involved in the area functional we performed a change of variables \eqref{eq:barredvariables} which brought the integrals to a form matching the vacuum integrals \eqref{eq:Iiinbarredcoords}. To set the limits of integration over $\bar\rho$ we need to express the FG cut-off in the new coordinates as  $\bar{\rho}_c(\bar\theta,\bar\alpha,\varepsilon)$.

The first step is to express $\{ \rho,\theta,\alpha \}$ coordinates in terms of $\{ \bar\rho,\bar\theta, \bar\alpha \}$. Combining (\ref{eq:changecoordinates}, \ref{eq:barredvariables}, \ref{eq:barredvariables2}) we can write the change of variables as
\begin{align}\label{changev}
 \sqrt{\rho^2+1} \cos\theta  &=  y_i \sqrt{\bar \rho^2+1} \cos\bar \theta \nonumber \\
 \rho \sin\theta \cos(\alpha) &= y_i \bar \rho \sin\bar \theta \cos\bar \alpha+ r_i^2 \cos^2\beta_i\nonumber\\
 \rho \sin\theta \sin(\alpha) &= y_i  \bar \rho \sin\bar \theta \sin\bar \alpha+ r_i^2 \sin^2\beta_i
\end{align}
where we have defined $x_{i1} = r_i \cos\beta_i$ and $x_{i2} = r_i \sin\beta_i$. We begin with solving the first equation in terms of $\rho$. Then, we combine the last two equations to eliminate $\alpha$ and we substitute $\rho$. This gives an equation for $\sin\theta$ in terms of the barred variables:
\begin{align}
\sin^4\theta+ A\, \sin^2\theta+B=0
\end{align}
with
\begin{align}
A&= -1 + r_i^2  + 2 \bar\rho\, r_i \,y_i\cos \left(\bar\alpha-\beta_i \right) \sin\bar\theta+ \frac{y_i^2 \left(1+\cos 2\bar\theta + 2 \bar\rho^2 \right)}{2} \nonumber\\
B&=-r_i^2- 2 \bar\rho\, r_i \,y_i \cos \left(\bar\alpha-\beta_i \right) \sin\bar\theta+\frac{ \bar\rho^2\,y_i^2\left( \cos 2\bar\theta-1\right)}{2}
\end{align}
Since $\theta\in[0,\pi/2]$ we choose the solution for which $\sin\theta$ is real and positive. We get the rest by plugging this solution into the equations \eqref{changev} . Specifically, $\rho$ is found by plugging $\sin\theta$ into the first equation while $\sin\bar\alpha$ and $\cos\bar\alpha$ are found using the other two equations. Since we only need the asymptotic behavior we give the results expanded at large $\bar\rho$:
\begin{align}\label{unbarredExpa}
\rho^2 &= y_i^2 \bar \rho^2 +2 r_i y_i \cos (\bar\alpha+\beta_i) \sin\bar\theta \bar\rho + \frac{1}{2} \left( y_i^2+ 2 r_i^2 -1 + (y_i^2-1) \cos 2\bar\theta \right) + O \left(\frac{1}{\bar\rho}\right) \nonumber\\
\sin^2 \theta &= \sin^2 \bar\theta + \frac{2 r_i \cos (\bar\alpha+\beta_i) \cos^2\bar \theta\sin\bar\theta}{y_i \bar\rho}\nonumber\\
& \phantom{=\ } + \frac{\cos^2\bar\theta \left( 1-y_i^2+(y_i^2+2r_i^2-1)\cos 2\bar\theta -4 r_i^2 \cos (2\bar\alpha+2\beta_i)\sin^2\bar\theta\right)}{2 y_i^2\bar \rho^2}+ O \left(\frac{1}{\bar\rho^3}\right)\nonumber\\
\sin\alpha &= \sin\bar\alpha+ \frac{r_i \csc\bar\theta (\cos\beta_i - \cos(\bar\alpha + \beta_i)\sin\bar\alpha)}{y_i \bar\rho} \nonumber\\
& \phantom{=\ } + \frac{r_i^2 \csc^2 \bar\theta \left( -4\cos\beta_i \cos(\bar\alpha+\beta_i)+\sin\bar\alpha+3\cos(2\bar\alpha +2\beta_i) \sin\bar\alpha\right)}{4 y_i^2 \bar\rho^2} + O \left(\frac{1}{\bar\rho^3}\right) \nonumber\\
\cos\alpha &= \cos\bar\alpha- \frac{r_i \csc\bar\theta (\cos\beta_i +\cos(2\bar\alpha + \beta_i)-2\sin\beta_i)}{2 y_i \bar\rho} \nonumber\\
& \phantom{=\ } + \frac{r_i^2 \csc^2 \bar\theta \left( -4\sin\beta_i \cos(\bar\alpha+\beta_i)+\cos\bar\alpha+3\cos(2\bar\alpha +2\beta_i) \cos\bar\alpha\right)}{4 y_i^2 \bar\rho^2} + O \left(\frac{1}{\bar\rho^3}\right)
\end{align}

To find the cut-off  $\bar{\rho}_c(\bar\theta,\bar\alpha,\varepsilon)$ we substitute \eqref{unbarredExpa} in the expression for the FG coordinate $u$, which can be found in \eqref{eq:FGMap}, to get $u$ in terms of the barred coordinates.
\begin{align}
u&= \frac{1}{y_i \bar\rho} - \frac{r_i \cos(\bar\alpha+\beta_i) \sin\bar\theta}{y_i^2 \bar\rho^2} - \frac{1}{8 y_i^3 \bar\rho^3} \Big[ 1 + 4 r_i^2 + 2y_i^2 + 2(y_i^2-1) \cos 2\bar\theta \nonumber\\
& \phantom{=\ }  + 2 m_{220} + 2 m_{202} - m_{400} - 3 \sin^2\bar\theta \left( 1 + 2r_i^2 + 2r_i^2 \cos(2\bar\alpha+2\beta_i) \right. \nonumber\\
& \phantom{=\ } \left. + 2 m_{220} + 2 m_{202} - m_{400}+4\sin 2\bar\alpha\, m_{211} + 2\cos 2\bar\alpha (m_{220}-m_{202}) \right) \Big] + O \left(\frac{1}{\bar\rho^4}\right) 
\end{align}
Solving this asymptotically for $\bar\rho$ and setting $u=\varepsilon$ we find the cut-off surface in barred coordinates.
\begin{align}
\bar{\rho}_c(\varepsilon,\bar\theta,\bar\alpha) &= \frac{1}{y_i \varepsilon} - \frac{r_i \cos(\bar\alpha+\beta_i) \sin\bar\theta}{y_i} + \frac{1}{8 y_i} \Big[ -1-4r_i^2-2y_i^2- 2(y_i^2-1)\cos2\bar\theta \nonumber\\
& \phantom{=\ } - 2m_{220}-2m_{202}+m_{400} + \sin^2\bar\theta \left(3+2r_i^2+2r_i^2 \cos(2\bar\alpha+2\beta_i) \right. \nonumber\\
& \phantom{=\ } \left.+6m_{220}+6m_{202}-3m_{400}+12\sin 2\bar\alpha \,m_{211}+6\cos 2\bar\alpha (m_{220}-m_{202})\right) \Big] \varepsilon + O \left( \varepsilon^2\right) 
\end{align}

%%%%%%%%%%%%%%%%%
\section{Coordinate systems and maps}
\setcounter{equation}{0}
\label{appCoordMaps}
%%%%%%%%%%%%%%%%%
In this section we collect useful formulae for the various coordinate systems and their maps along with information about our setup in these systems. In particular we relate $AdS_3\times S^1$ to $S^1 \times H^3$ with an intermediate transformation to $\mathbb{R}^4$. In the latter space the picture of our setup becomes more clear (see figure~\ref{fig:R4-setup}). 

The metrics on the 4D Euclidean spaces we consider are the following:
\begin{align}\label{eq:4dmetrics}
AdS_3 \times S^1& &  &ds^2_{AdS_3 \times S^1} = \frac{dt^2+dl^2+dz^2}{z^2} + d\psi^2 \nonumber\\
\textrm{spherical}&  & &ds^2_{\mathbb{R}^4} =  dt^2+dx^2 +x^2\left(d\vartheta^2+\sin^2\vartheta\, d\psi^2\right) \nonumber\\
\textrm{hyperboloid}& & &ds^2_{S^1\times H^3} = d\tau^2 + R^2 \left( d\rho^2 +\sinh^2\rho\left( d\vartheta^2+\sin^2\vartheta\, d\psi^2\right)\right)
\end{align}
They are conformally related to each other as follows:
\begin{align}
ds^2_{AdS_3 \times S^1} = z^{-2}\, ds^2_{\mathbb{R}^4}, \quad ds^2_{\mathbb{R}^4} = \bar{\Omega}^2\, ds^2_{S^1\times H^3}, \quad ds^2_{AdS_3 \times S^1} = \Omega^2\, ds^2_{S^1\times H^3}
\end{align}
where
\begin{align}\label{eq:OmegaConformalfactor}
\bar{\Omega} = \left( \cosh\rho+ \cos(\tau/R)\right)^{-1} ,\quad \Omega=\left(R \sinh\rho\sin\vartheta\right)^{-1}
\end{align}
The coordinate maps corresponding to these three transformations are given by 
\begin{align}\label{eq:4dCoordinateMaps}
AdS_3 \times S^1 \;\textrm{to}\; \textrm{spherical:}& & &l= x \cos\vartheta, \;\; z = x\sin\vartheta \nonumber\\
\textrm{spherical}  \;\textrm{to}\; \textrm{hyperboloid:}& & &t= R\; \bar\Omega \sin(\tau/R), \;\; x= R\; \bar\Omega \sinh\rho \nonumber\\
AdS_3 \times S^1 \;\textrm{to}\; \textrm{hyperboloid:}& & &t=R\; \bar\Omega \sin(\tau/R), \;\; l=R\; \bar\Omega \sinh\rho \cos\vartheta,\;\; z = R\; \bar\Omega \sinh\rho \sin\vartheta
\end{align}
where the last transformation comes from combining the first two. 

For easy reference we quote the location $\Sigma$ of the surface defect and the location $\partial \cal{A}$ of the entangling surface in the various coordinate charts: 
\[
\begin{array}{rcc}
& \Sigma & \partial \cal{A}\\
AdS_3\times S^1  & \textrm{fills\ } t, \textrm{fills\ } l, z=0 & t=0, l^2+z^2=R^2\\
\textrm{spherical} &\textrm{fills\ } t, \textrm{fills\ } x, \vartheta=0, \pi & t=0, x=R \\
\textrm{hyperboloid} &\textrm{fills\ } \tau, \textrm{fills\ } \rho, \vartheta=0,\pi &  \rho\to\infty \\
\end{array}
\]
It can be seen, in all coordinate charts, that the surface defect intersects the entangling surface exactly at two points.

\section{Asymptotic expansion comparison with \cite{Drukker:2008wr}}
\setcounter{equation}{0}
\label{appstress}
%%%%%%%%%%%%%%%%%%%%
For calculating holographic observables one has to expand the supergravity solution in an asymptotic form. In this section we quote the way the asymptotic expansion was performed in \cite{Drukker:2008wr} and compare with ours. 

Defining $\Phi=f/y$ the equation for $f$, \eqref{fdiffeq}, can be written as the six-dimensional Laplace equation  for $\Phi$ with $SO(4)$ invariant sources. In \cite{Drukker:2008wr} the authors write $\Phi$ as the vacuum part and a deviation:
\begin{align}
\Phi=\Phi^{(0)}+\Delta\Phi
\end{align}
Then, they expand the deviation  $\Delta\Phi$ in $SO(4)$-invariant spherical harmonics. The coefficients of this expansion are denoted by $\Delta\Phi_{\Delta,k}$, where $\Delta,k$ are eigenvalues characterizing the spherical harmonics (for more details on the spherical harmonics see appendix~A in \cite{Drukker:2008wr}).

As an example, we quote their result for the one-point function of the stress tensor which was found using holography:
\begin{align}\label{stressTensorGomis}
\langle T_{\mu\nu}\rangle_\Sigma \;dx^\mu dx^\nu = \frac{N^2}{2\pi^2} \left(  \frac{1}{16} - \frac{1}{12\sqrt{3}}\,\Delta \Phi_{2,0}  \right)  \left(ds^2_{AdS_3} - 3 \, d\psi^2\right)
\end{align}
One can see that this matches (\ref{stressTensorForm2}, \ref{hResult}), when a definition for $\Delta \Phi_{2,0}$ is given in terms of the moments. For completeness we give all the coefficients corresponding to spherical harmonics with eigenvalue $\Delta=2$ in terms of the moments:
\begin{align}\label{eq:sphHarmonicsExpa}
\Delta \Phi_{2,0} &=  4\sqrt{3} \left( m_{220}+ m_{202} +\frac{1-m_{400}}{2} \right) \nonumber\\
\Delta \Phi_{2,\pm2} &= 6 e^{\mp 2 i \psi} \left(m_{220}-m_{202}\pm 2 i m_{211}\right)
 \end{align}

%%%%%%%%%%%%%%%%%%%%
\section{Holographic expectation value}
\setcounter{equation}{0}
\label{appast}
In this appendix we compute the integrals involved in the expectation value of the surface defect \eqref{expectationValueOnShell}. Specifically, these are the bulk contribution given in \eqref{actionElectric} and the Gibbons-Hawking term in \eqref{GHterm}.

\subsection{Bulk term}\label{appastBulk}
Let us start with the evaluation of the bulk term. The method described in \cite{Giddings:2001yu,DeWolfe:2002nn} led us to \eqref{actionElectricIntegrand} the integrand of which we expressed as \eqref{bulkElectricSimplified}. We begin with carrying out the integration over $AdS_3$, $S^3$ and $S^1$, which is trivial. Then, the bulk term can be expressed in terms of two integrals over the base space $X$:
\begin{align}
I_{\textrm{bulk}}=-\frac{4}{\kappa^2}\,\textrm{Vol}\left(AdS_3\right) \textrm{Vol}\left(S^3\right) \textrm{Vol}\left(S^1\right) \left[ -\frac{1}{2}\, J_1 + J_2 \right]
\end{align}
where we have defined:
\begin{align}
J_1 &= \int_X dx_1\, dx_2\, dy\, f y \\
J_2 &= \int_X dx_1\, dx_2\, dy\, \partial_I u_I
\end{align}

Making use of the integral \eqref{eq:I} appearing in the entanglement entropy calculation we can write
\begin{align}
J_1 &= \int_X dx_1\, dx_2\, dy\, \left[\left(f - \frac{1}{2}\right) y + \frac{1}{2}\, y\right]\\
  &= \frac{\pi}{4\varepsilon^2}+\frac{\pi}{24} \, \left[1 - 4\left(m_{220}+m_{202}+m_{400}\right)\right]+ \frac{1}{2}\, \int_X dx_1\, dx_2\, dy\, y
\end{align}
 where we have dropped terms that vanish as $\varepsilon\to 0$.  The  integral in the second line can be evaluated directly by changing to $\{\rho,\theta,\alpha\}$ coordinates (the relevant map is given in \eqref{eq:changecoordinates}):
\begin{align}
\int_X dx_1\, dx_2\, dy\, y
 &= \int d\rho\, d\theta\, d\alpha\, \rho \left(\rho^2+\sin^2\theta\right)\cos\theta\sin\theta \nonumber\\
  &= \int_0^{\pi/2}  d\theta \int_0^{2\pi} d\alpha\, \left. \frac{1}{4}\, \rho ^2  \left(
 \rho ^2 +2  \sin ^2\theta \right) \cos \theta\sin \theta   \right|_0^{\rho_c(\varepsilon,\theta,\alpha)} \nonumber\\
 &= \frac{\pi}{4\varepsilon^4} +  \frac{\pi}{16\varepsilon^2}\left(1+2m_{220}+2m_{202}-m_{400}\right) + Y_1
 \end{align}
where the term $Y_1$ reads:
\begin{align}
 Y_1&\equiv \frac{\pi}{768}\left[ -7  +12 m_{220}  +12 m_{202}-6m_{400}\right. \nonumber\\
  &\phantom{= \ }-288 \left(m_{220}^2 +m_{202}^2\right)+144 m_{220}m_{202}-720 m_{211}^2+108 \left(m_{220}+m_{202}\right)m_{400}-27 m_{400}^2 \nonumber\\
   &\phantom{= \ }\left. +48 \left(m_{240}+ m_{204}\right)    + 96 \left(m_{222} -m_{402}-m_{420}\right)+16m_{600} \right]
 \end{align}
 
 \begin{figure}[!t]
  \centering
  \includegraphics[width=90mm]{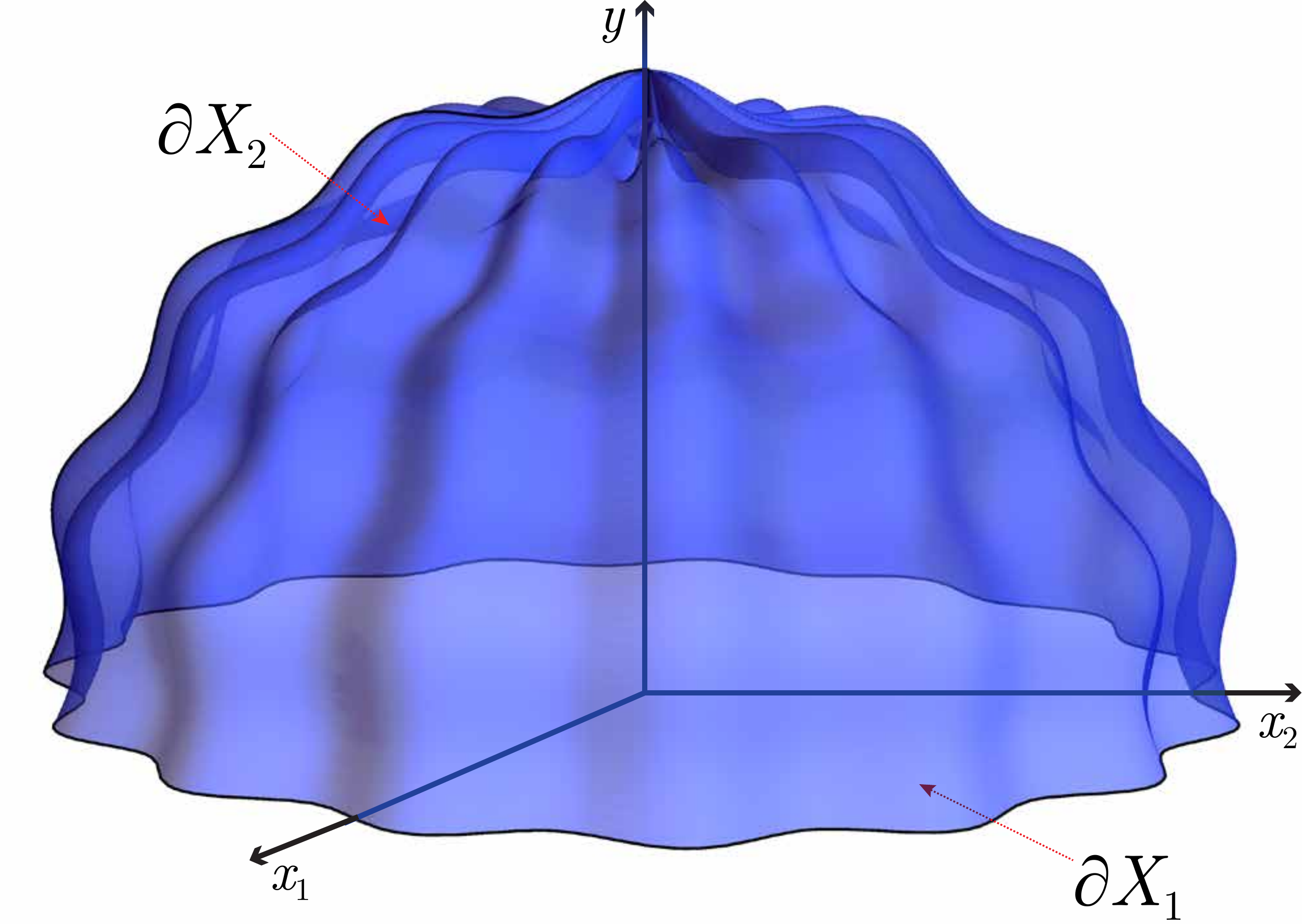}
  \caption{The base space $X$ boundary components: the blue wiggled dome noted as $\partial X_2$ is the large $\rho$ cut-off and the lowest flat surface noted as $\partial X_1$ is the boundary at the $x_1,x_2$ plane.}
 \label{fig:dome}
\end{figure}

Next we evaluate $J_2$ by turning it into an integral over the boundary of  $X$.  Switching to covariant notation in which $g_{IJ}$ is a metric on $X$ we have
\begin{align}
J_2 &= \int_X d^3x\, \sqrt{g}\, \nabla_I\, u^I \\
 &= \int_{\partial X} d^2x\, \sqrt{\gamma}\, n_I\, u^I
\end{align}
where $n$ is the outward-pointing unit normal vector and $\gamma$ the induced metric on $\partial X$.    This surface consists of two components (see figure~\ref{fig:dome}):
\begin{align}
\partial X_1 &= \left\{\left.\left(x_1,x_2,y\right) \right| y=0, x_1^2+x_2^2 \leq \rho_c(\varepsilon,\pi/2,\alpha)^2 \right\}\\
\partial X_2 &= \left\{\left.\left(\rho,\theta,\alpha\right) \right| \rho = \rho_c(\varepsilon,\theta,\alpha), \theta\in [0,\pi/2], \alpha\in [0,2\pi] \right\}
\end{align}
The contribution to $J_2$ from $\partial X_1$ vanishes. This can be easily seen by expanding \eqref{totalDerivativeVector} for small $y$ and take the $y\to0$ limit. For the remaining contribution we work in $\{\rho,\theta,\alpha\}$ coordinates. The metric on $X$ is 
\begin{equation}
ds^2_X =\frac{\rho^2 + \sin^2\theta}{\rho^2+1}\, d\rho^2 + \left(\rho^2 + \sin^2\theta\right) d\theta^2 + \rho^2  \sin^2\theta\,d\alpha^2
\end{equation}
The unit  vector normal to the surface $\rho - \rho_c(\varepsilon,\theta,\alpha)=0$ has the following components in this chart:
\begin{gather}
n_\rho = \frac{1}{\cal D}, \quad n_{\theta} = -  \frac{\partial_{\theta}\rho_c(\varepsilon,\theta,\alpha)}{\cal D}, \quad n_\alpha = - \frac{\partial_{\alpha}\rho_c(\varepsilon,\theta,\alpha)}{\cal D}\\
{\cal D} \equiv  \sqrt{\frac{[\partial_{\alpha}\rho_c(\varepsilon,\theta,\alpha)]^2}{\rho^2  \sin^2\theta}+\frac{\rho^2 + 1+[\partial_{\theta}\rho_c(\varepsilon,\theta,\alpha)]^2}{\rho^2 + \sin^2\theta}} \label{eq:defineD}
\end{gather}
The induced metric and pullback components are given by
\begin{equation}
\gamma_{ab} = g_{IJ}\,e^I_a \, e^J_b \quad \textrm{with}\quad e^I_a = \left(\begin{array}{cc}
\partial_{\theta}\rho_c(\varepsilon,\theta,\alpha) &  \partial_{\alpha}\rho_c(\varepsilon,\theta,\alpha) \\
1 & 0 \\
0 & 1
\end{array}\right)
\end{equation}
where $a\in \{\theta,\alpha\}$.   We are now ready to evaluate $J_2$:
\begin{align}
J_2 &= \int_0^{\pi/2}  d\theta \int_0^{2\pi} d\alpha\, \sqrt{\gamma}\ \frac{y^3}{4 \left(4f^2-1\right)}\, g^{IJ}\, n_I\, \partial_J f\nonumber\\
&=- \frac{\pi}{16\varepsilon^4} +  \frac{\pi}{64\varepsilon^2}\left(1+2m_{220}+2m_{202}-m_{400}\right) +Y_2
\end{align}
where  
\begin{align}
 Y_2&\equiv \frac{\pi}{3072}\left[ -51  -100 m_{220}  -100 m_{202}+50m_{400}\right. \nonumber\\
  &\phantom{= \ }+72 \left(m_{220}^2 +m_{202}^2\right)+144 m_{211}^2-36 \left(m_{220}+m_{202}\right)m_{400}+9 m_{400}^2 \nonumber\\
   &\phantom{= \ }\left. +48 \left(m_{240}+ m_{204}\right)    + 96 \left(m_{222} -m_{402}-m_{420}\right)+16m_{600} \right]
\end{align}
  
Putting everything together we get
\begin{align}\label{bulkActionResult}
I_\textrm{bulk}= \frac{\pi}{2 \kappa^2} \,\textrm{Vol}\left(AdS_3\right) \textrm{Vol}\left(S^3\right) \textrm{Vol}\left(S^1\right)\left[ \frac{1}{\varepsilon^4}+\frac{1}{\varepsilon^2}+  \frac{3}{8} -  m_{400}+ \frac{2}{\pi} \left(Y_1-4Y_2\right)  \right]
% \frac{1}{128} \left(48- 128 m_{400}-\Delta\Phi_{2,k}\Delta\Phi_{2,-k}\right) \right]
\end{align}
Plugging in the explicit expressions for $Y_1$ and $Y_2$ we notice that the moments of weight six drop out. The result is given in (\ref{bulkActionResult2}, \ref{eq:Finmoments}).

\subsection{Gibbons-Hawking term}\label{appastGH}
To compute the Gibbons-Hawking term \eqref{GHterm} we use a similar method to that used in the previous subsection for the total derivative on $X$, but now in the full ten-dimensional spacetime.  The unit vector normal  to the surface $\rho - \rho_c(\varepsilon,\theta,\alpha)=0$ has the following non-trivial components
\begin{equation}
n_\rho = \frac{1}{{\cal D}\, \sqrt{\frac{2y}{\sqrt{4f^2-1}}}}, \quad n_{\theta} = -  \frac{\partial_{\theta}\rho_c(\varepsilon,\theta,\alpha)}{{\cal D}\, \sqrt{\frac{2y}{\sqrt{4f^2-1}}}}, \quad n_\alpha = - \frac{\partial_{\alpha}\rho_c(\varepsilon,\theta,\alpha)}{{\cal D}\, \sqrt{\frac{2y}{\sqrt{4f^2-1}}}}
\end{equation}
where $\cal D$ is defined in \eqref{eq:defineD}. The induced metric and non-trivial pullback components are given by
\begin{gather}
\gamma_{ab} = g_{MN}\,e^M_a \, e^N_b \\
e^\rho_\theta = \partial_{\theta}\rho_c(\varepsilon,\theta,\alpha),\quad e^\rho_\alpha=  \partial_{\alpha}\rho_c(\varepsilon,\theta,\alpha), \quad e^a_b = \delta^a_b
\end{gather}
where now $a$ runs over all coordinates except $\rho$.  The extrinsic curvature can be computed from the Lie derivative along $n$:
    \begin{align}
 K_{ab} &= \frac{1}{2} \left({\cal L}_n\, g\right)_{MN} e^M_a \, e^N_b \\
  &= \frac{1}{2} \left(n^P\, \partial_P g_{MN} +g_{PN}\, \partial_M n^P +g_{MP}\, \partial_N n^P\right)e^M_a \, e^N_b 
  \end{align}
 and its trace is simply $K\equiv \gamma^{ab}\, K_{ab}$ (whose small $\varepsilon$ expansion leads with order 4).  The result is 
 \begin{equation}\label{eq:GHresult}
 I_{\textrm{GH}} = \frac{\pi}{2\kappa^2}\,\textrm{Vol}\left(AdS_3\right) \textrm{Vol}\left(S^3\right) \textrm{Vol}\left(S^1\right) \left(\frac{4}{\varepsilon^4} +  \frac{1}{\varepsilon^2} \right)
\end{equation}
The moments appearing in the boundary integrand drop out when the integration over the angles $\{\theta,\alpha\}$ is performed.

Note that there is in principle a contribution from the other component of the boundary at $y=0$, but again this vanishes. Specifically, expanding the Gibbons-Hawking integrand for  small $y$ we get $\sqrt\gamma\,  K=O\left(y^2\right)$ which vanishes in the $y\to0$ limit.

%%%%%%%%%%%%%%%%%%%%

\newpage

%%%%%%%%%%%%%%%%%%%%%%%%%
 %%% Bibliography
 %%%%%%%%%%%%%%%%%%%%%%%%%

\providecommand{\href}[2]{#2}\begingroup\raggedright\endgroup

\end{document}